# Unveiling ferroelectricity in wafer-scale beta-phase $In_2Se_3$

*Kyungmin Yang[a,+], Hyunmin Cho[a,+], Jihwan Jeon[b,c,+], Tae Hyeon Park[b,c], Seojeong Ro[c], Hye Jung Kim[b], Jung-Hoon Lee[d,e], Jaewon Wang[a], Hyeon Suk Shin[b,c,f], Young-Min Kim[b,c], Chang Woo Myung[b,c], Roy H. Olsson III[a], Seunguk Song[b,c,*], Deep Jariwala[a,*]*

[a] Department of Electrical and Systems Engineering, University of Pennsylvania, Philadelphia, PA 19104, USA

[b] Center for 2D Quantum Heterostructures (2DQH), Institute for Basic Science (IBS), Suwon, 16419, Republic of Korea

[c] Department of Energy Science, Sungkyunkwan University (SKKU), Suwon, 16419, Republic of Korea

[d] Computational Science Research Center, Korea Institute of Science and Technology (KIST), Seoul, 02792, Republic of Korea

[e] KU-KIST Graduate School of Converging Science and Technology, Korea University, Seoul, 02841, Republic of Korea

[f] Department of Chemistry, Sungkyunkwan University (SKKU), Suwon, 16419, Republic of Korea

[+] These authors contributed equally

* E-mail: seunguk@skku.edu, dmj@engineering.upenn.edu

## Abstract

Pushing ferroelectricity to the ultimate thickness limit is a central challenge for memory hardware, because conventional three-dimensional ferroelectrics degrade through depolarization and interfacial effects when thinned. Here we identify β-$In_2Se_3$, a van der Waals semiconductor regarded as centrosymmetric, as a host of bistable polar order arising from a picometre-scale distortion within each quintuple layer. The polar state has out-of-plane and in-plane components, is switched by a vertical electric field, and persists to the bilayer limit (~1.9 nm) with an estimated coercive field of 0.9 MV/cm and an apparent remanent polarization of 11.0 μC/cm$^2$. Films grown by metal-organic chemical vapor deposition on 2-inch wafers yield capacitor arrays with a switching-voltage spread below 5% across 100 devices, an endurance of $10^7$ cycles and a projected ten-year retention. Because the polar state coexists with a semiconducting gap, one film serves as both the switching layer of a capacitor and the channel of a ferroelectric semiconductor transistor.

## Introduction

Ferroelectric materials possess a spontaneous electric polarization that can be reversibly switched by an external electric field, making them attractive for non-volatile memory, logic and emerging computing technologies[1,2]. Extending ferroelectricity to the atomic scale is particularly important for continued device scaling, as switchable polarization can enable low-power information storage and processing within ultrathin active layers. However, conventional three-dimensional (3D) ferroelectrics become increasingly difficult to stabilize as their thickness is reduced to the nanometer regime. Strong depolarization fields, interfacial effects and structural instabilities can suppress or degrade ferroelectricity in ultrathin films[3-5].

Two-dimensional (2D) materials provide an alternative platform for realizing ferroelectricity at reduced dimensions[6]. Their atomically thin nature and van der Waals bonding can alleviate several limitations associated with conventional ferroelectric scaling, while also enabling heterogeneous integration without strict lattice-matching requirements. Among 2D ferroelectrics, indium selenides have attracted particular attention because they can exhibit ferroelectricity while retaining semiconducting characteristics[7]. The coexistence of switchable polarization and semiconducting transport offers opportunities for compact and energy-efficient devices for in-memory computing, in-sensor computing, neuromorphic electronics and optoelectronics[8-10].

In practice, indium selenides remain limited by challenges associated with phase and stacking control. In γ-InSe, which exhibits sliding ferroelectricity[11], the polarization strongly depends on the relative stacking configuration between adjacent layers, making deterministic control of the ferroelectric state difficult over large areas. In α-$In_2Se_3$, which exhibits spontaneous polarization[12], structural phase transformations can occur within the back-end-of-line processing temperature window[13], limiting its scalability for large-area synthesis and monolithic device integration. On the other hand, the β phase is the end product of this transformation and the polymorph most readily obtained by vapour-phase growth at the temperatures required for wafer-scale films[14,15] (a κ polymorph has also been grown recently, but it is not ferroelectric[16]), which makes an understanding of its polar properties a prerequisite for scalable indium selenide electronics. A distorted β′ polymorph with a nanometre-period stripe superstructure and in-plane antipolar order is known[17,18], but it carries no switchable out-of-plane polarization. The β phase itself has been classified as paraelectric on structural grounds, and the hysteresis observed in β-$In_2Se_3$ transistors has been attributed either to charge trapping or, without a structural model of a polar state, to polarization switching[14,19]. Therefore, whether β-$In_2Se_3$ hosts a switchable polarization, in particular an out-of-plane one, has remained an open question (Supplementary Table 3). Switchable polarization has recently been reported in large-area indium selenide films with β-like layers grown by molecular beam epitaxy, but without a field-reversible atomic displacement, an identified polar ground state, or a quantified remanent polarization and coercive field[20,21].

In this study, we report that the nominally centrosymmetric β phase of $In_2Se_3$ is intrinsically ferroelectric. In contrast to α-$In_2Se_3$, in which polarization requires a shift of the central Se layer by ~0.4 Å, and to β′, in which the distortion is confined to the plane, the polar state identified here is set by a field-reversible displacement of a few picometres. This displacement carries both out-of-plane and in-plane components and can be written with a vertical field. Using wafer-scale-grown β-$In_2Se_3$, scanning transmission electron microscopy (STEM) reveals bias-dependent atomic displacements that vary with the direction of the applied electric field. First-principles calculations further show that the distorted β-$In_2Se_3$ structure supports two energetically stable polarization states and exhibits a characteristic double-well potential, consistent with bistable ferroelectric switching. Experimentally, out-of-plane polarization switching is observed in capacitors, and lateral devices show a hysteretic response that is consistent with switching of the in-plane component. The out-of-plane response persists down to the bilayer limit (~1.9 nm), with an estimated coercive field of 0.9 MV/cm and an apparent remanent polarization of 11.0 $\mu C/cm^2$ measured for the bilayer. The same wafer-scale films yield uniform capacitor arrays and ferroelectric semiconductor field-effect transistors. These findings reveal symmetry-breaking lattice distortion in β-$In_2Se_3$ as a distinct route to ferroelectricity and establish a scalable platform for atomically thin, low-power non-volatile electronics.

## Results and Discussion

β-phase $In_2Se_3$ has long been classified as paraelectric on the basis of its assumed centrosymmetric structure[14]. It consists of Se-In-Se-In-Se quintuple layers (QLs) stacked along the c axis and separated by van der Waals gaps, with a quintuple-layer stacking period $d_{\mathrm{inter}}$ = 0.9443 nm and an in-plane atomic spacing of 0.4000 nm (Figure 1a). Unlike the ferroelectric α phase, in which one In sublattice is tetrahedrally coordinated, all In atoms of the β phase occupy octahedral sites. In the idealized centrosymmetric structure (space group $R\bar{3}m$), the central Se layer ($Se_C$) sits exactly midway between the two In layers, consistent with the conventional classification of the β phase as paraelectric[14]. However, this classification does not rule out a small atomic distortion (as depicted in atomic schematics in Figure 1b) that may not have been resolved by conventional diffraction. We therefore looked for such a distortion directly, at the level of individual atomic columns. For this purpose, we first grew wafer-scale β-$In_2Se_3$ films by MOCVD on 2-inch c-plane sapphire (Figure 1c; Methods). Raman spectroscopy, atomic force microscopy (AFM) and X-ray photoelectron spectroscopy (XPS) were consistent with the β phase and showed no detectable oxidation-related bonding signals (Supplementary Figures 1-2); the XPS composition (Se/In ≈ 1.47) is used only as a consistency check, because the accuracy of XPS quantification is limited to about ±2%. The phase was identified directly from the atomic-resolution cross-sectional STEM images, in which the quintuple-layer stacking sequence and the atomic-column arrangement match the simulated image of the β structure (Figure 1d and Supplementary Figures 3 and 4), and the same images confirmed the epitaxial relationship between the film and the sapphire substrate (Supplementary Figure 3).

Note that the β′ polymorph is defined by a sign-alternating in-plane displacement of the central Se layer with a period of a few nanometres[17,18], which would appear as a periodic modulation in cross-sectional images. Neither cross-section shows such a modulation and the in-plane displacements resolved below are uniform in sign over the imaged regions. These observations provide no evidence of the β′ superstructure, and we therefore identify the films as β.

We then fabricated capacitor devices on these films, set them to the Program or Erase states (P/E states) by applying out-of-plane electric fields (Methods), and imaged their cross-sections along the $[11\bar{2}0]$ zone axis by high-angle annular dark-field STEM (HAADF-STEM) (Figure 1d). To compare the two P/E states using a local measure of structural asymmetry, we used as the observable the difference in length, Δ, between the two In–$Se_C$ bonds within the same QL, which is zero for a centrosymmetric QL and changes sign when the distortion reverses. Using a qSTEM-simulated image as the reference (Supplementary Figure 4), we constructed in-plane and out-of-plane projected atomic distance (PAD) maps for each state from the atomic-column positions (Figure 1d and Supplementary Figures 5-8). For each PAD component, the fast-scan direction was set parallel to the measured component to minimize scan-distortion artifacts (Methods). In the out-of-plane direction, the two In–$Se_C$ separations in a QL differed by 6.0 ± 0.9 pm after Program and by 4.8 ± 1.2 pm after Erase, with opposite signs (mean ± s.e.m. over 206 and 209 QL sequences, respectively; Supplementary Figures 5 and 6). These differences correspond to a displacement of $Se_C$ from the In–In midpoint of +3.0 ± 0.4 pm after Program and −2.4 ± 0.6 pm after Erase, where the uncertainty is the statistical standard error and systematic scan distortion is suppressed by the fast-scan geometry described in Methods. The displacement is small, but it is not random. Atomic-column positions in HAADF-STEM images can be determined with a precision of several picometres, and below 1 pm when an image series is registered and averaged[22,23]; averaged over more than 200 QLs, the displacements measured here have standard errors of only 0.4–0.6 pm, and the 5.4 pm difference between the two states lies well outside the measurement error.

Displacement vector maps constructed from the same images, in which each arrow indicates the direction of the displacement of the In–Se framework relative to $Se_C$ within one QL (Figure 1e), show the vectors aligned in one direction after Program and in the opposite direction after Erase, uniformly across the imaged region of ≈8.6 × 8.6 $nm^2$. Because each component was measured from an image whose fast-scan direction was parallel to it, and because the two states were imaged in independent lamellae, scan distortion cannot account for the opposite signs. In-plane, the Program lamella shows a $Se_C$ offset of ≈9 pm whereas the Erase lamella shows ≈1 pm (Supplementary Figures 7 and 8), which is consistent with the two lamellae sampling different in-plane variants of the polar structure. The out-of-plane component is common to all variants and is therefore the quantity compared between the two states. A bond-length asymmetry around the central Se has also been noted in MBE-grown $In_2Se_3$ films[20], where it was assigned to a distinct polymorph on the basis of STEM, second-harmonic generation and transistor hysteresis, but without poling-dependent reversal, a polar ground-state structure or a quantified polarization.

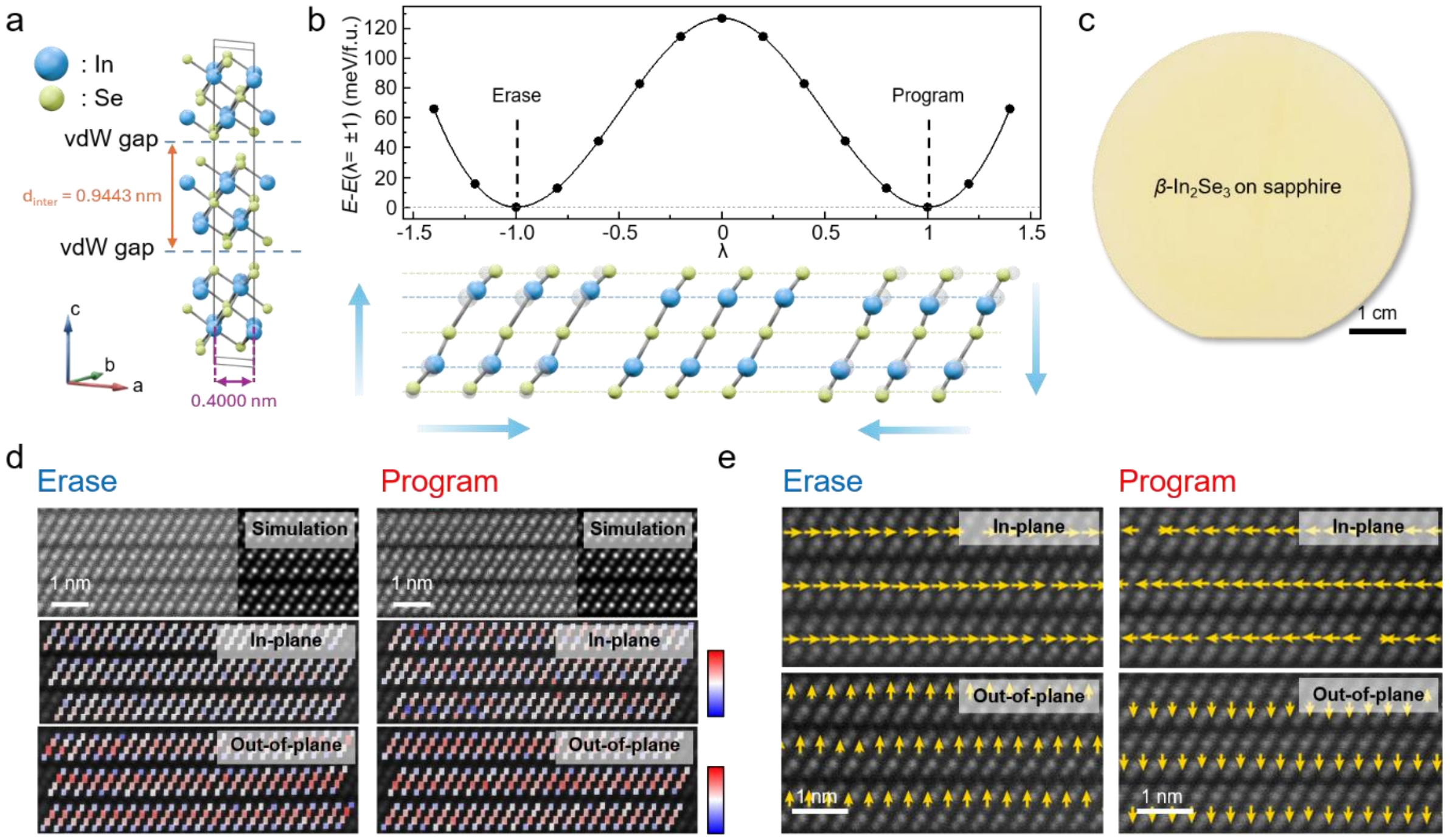

**Figure 1. Atomic-scale in-plane and out-of-plane mapping of the two polarization states in wafer-scale β-$In_2Se_3$.** (a) Crystal structure of β-$In_2Se_3$; blue and green spheres denote In and Se atoms, respectively. Quintuple layers (Se–In–Se–In–Se) are stacked along the c axis and separated by van der Waals gaps, with a quintuple-layer stacking period dinter = 0.9443 nm and an in-plane atomic spacing of 0.4000 nm. (b) (Top) Double-well potential energy landscape connecting the two polarization states. (Bottom) Corresponding atomic configurations. The Program and Erase states are related by inversion, with the In/Se framework displaced in opposite directions with respect to the central Se; semi-transparent spheres mark the atomic positions of the nonpolar reference structure, and blue arrows indicate the directions of the out-of-plane and in-plane polarization components. Atomic displacements are exaggerated by a factor of 10 for clarity. (c) Photograph of β-$In_2Se_3$ deposited on a 2-inch c-plane sapphire substrate. (d) Cross-sectional HAADF-STEM images of the device after Erase (left) and after Program (right), with qSTEM-simulated images of the centrosymmetric reference structure of β-$In_2Se_3$ along $[11\bar{2}0]$ shown as insets, together with the corresponding in-plane and out-of-plane projected atomic distance (PAD) maps. Color scales, 0.8–1.5 Å (in-plane) and 1.0–2.4 Å (out-of-plane). (e) Displacement vector maps obtained from the HAADF-STEM images in (d). Each arrow indicates the direction of the displacement of the In–Se framework relative to the central Se within one QL, for Erase (left) and Program (right). Arrow lengths scale nonlinearly with the displacement magnitude to aid visibility, as $L = L_{min} + (L_{max} - L_{min})\tanh(|u|/u_0)$, where $u$ is the displacement, $L_{min}$ = 18 pixels (1.8 Å), $L_{max}$ = 27 pixels (2.7 Å) and $u_0$ = 0.05 Å.

The polarity and energetic stability of the observed distorted structures were then examined by first-principles calculations. Relaxing the structure without symmetry constraints (HSE06+D4 at the experimentally measured lattice parameters; Methods) gives a polar equilibrium structure of $C_m$ symmetry in which $Se_C$ sits 4.5 pm off the midpoint between the two In layers along the c axis, comparable to the 3.0 and 2.4 pm measured by STEM. In this structure the In and Se sublattices are also displaced in the plane (Supplementary Figure 9). The energy along a path λ that connects this structure (λ = +1) to the oppositely polarized structure (λ = -1) through the centrosymmetric reference (λ = 0) has the double-well form characteristic of a ferroelectric (Figure 1b). The two minima are energetically degenerate, and the centrosymmetric structure lies 127 meV per formula unit above the two minima (Supplementary Figure 10). Along this path, the centrosymmetric structure assumed so far is thus not the ground state but the barrier between two polar states. The two out-of-plane polarizations at the minima have a magnitude of 1.87 μC/cm$^2$ and opposite signs, and the same polar mode carries an in-plane component of 21.1 μC/cm$^2$ that reverses along the same path (Supplementary Figure 10). Reversal of the dominant in-plane component by a field applied along the minor out-of-plane component is analogous to the perpendicular switching recently reported in layered $Bi_4Ti_3O_{12}$, in which the two components are tied to a single polar mode[24]. Both components are well defined along the whole path because the band gap (~1.35 eV; Supplementary Figure 11) remains open, which means that polarization reversal does not pass through a metallic state and that β-$In_2Se_3$ is a ferroelectric semiconductor in both states. The two minima are consistent with the P/E states observed by STEM. The out-of-plane displacement that separates them is almost ten times smaller than the ~0.4 Å shift of the central Se layer in its α-phase polymorph, yet the calculated out-of-plane polarization is comparable to that of α-$In_2Se_3$ (~0.11 eÅ per unit cell, that is, ~1.3 μC/cm$^2$)[25], indicating that β-$In_2Se_3$ stores a similar polarization with a much smaller atomic motion. The calculated values refer to the lattice polarization of the relaxed polar structure at zero field. They are smaller than the apparent remanent polarization of ~11–20 μC/cm$^2$ obtained from the capacitor measurements described below, and we attribute the difference to contributions that the calculation does not contain: in a semiconducting ferroelectric the charge switched in a capacitor includes the redistribution of screening carriers that accompanies polarization reversal, the measured charge retains part of the non-switching current that the PUND subtraction does not remove, and the displacement under the applied field may exceed its zero-field value. The measured values are therefore treated as upper bounds on the lattice polarization (see the discussion of Figure 2e). Two estimates that do not use the capacitor charge are close to the calculation: scaling the calculated polarization by the displacements measured by STEM (3.0 and 2.4 pm, versus 4.5 pm) gives 1.0–1.2 μC/cm$^2$, and the memory window of the $ZrO_2$-gated FeS-FET (4.31 V at ±5 V; Supplementary Table 2) corresponds to 1.5–1.9 μC/cm$^2$ for the gate capacitance of that stack. The lattice polarization of β-$In_2Se_3$ is therefore 1–2 μC/cm$^2$. In contrast to β′, in which the in-plane displacements order into alternating stripes, the distortion found here is uniform in sign over the imaged region, carries an out-of-plane component and reverses under a vertical field. These results support ferroelectricity in β-$In_2Se_3$, with the two polarization states distinguished by pm-scale relative atomic displacements within the QL, providing the structural basis for the electrical switching and retention described in the following sections.

Following the structural and theoretical analysis in Figure 1, we performed piezoresponse force microscopy (PFM) and electrical measurements to examine the out-of-plane and in-plane switching experimentally, with the corresponding sample preparation procedure illustrated in Supplementary Figure 12. Figure 2a shows the off-field PFM amplitude and phase hysteresis loops measured from a 22.7 nm-thick β-$In_2Se_3$ film on an Au bottom electrode. The thickness was determined by AFM, as shown in Supplementary Figure 13. The Au layer serves as the grounded bottom electrode, allowing an out-of-plane electric field to be applied between the conductive PFM tip and the bottom electrode, as illustrated in the inset of Figure 2a. The PFM response shows a butterfly-shaped amplitude loop and an approximately 180° phase reversal. In particular, the direction of the hysteresis loop indicated by the arrows in Figure 2a is consistent with ferroelectric polarization switching[26]. The local coercive voltages were identified from the minima in the PFM amplitude signal that coincided with the phase reversal[27]. Under positive tip bias, the out-of-plane electric field is directed toward the bottom electrode, and polarization switching from the up to the down state occurs at $V_C^+ \approx +17.4$ V (nominal $E_C \approx 7.7$ MV/cm). Conversely, switching from the down to the up state occurs at $V_C^- \approx -7.74$ V (nominal $E_C \approx -3.4$ MV/cm) under negative tip bias. The loop is shifted by ~5 V toward positive bias. Such offsets are common in tip-based PFM, in which the tip–surface contact, surface adsorbates and the dissimilar top (tip) and bottom (Au) boundaries shift the loop, and the offset is therefore not taken as an intrinsic imprint of the film[28]. In the capacitors, which use Au-supported films prepared in the same way, the switching onsets of the two polarities (Figure 2e) are offset by only ~2 V, and in the opposite field direction to the PFM loop, so the PFM offset is attributed to the tip contact rather than to a built-in field in the film. These nominal fields (obtained by dividing the tip bias by the film thickness) are upper bounds, because the tip-surface contact is ill defined and no top electrode screens the surface, so that tip-based switching voltages generally exceed the coercive voltages of capacitors[26,27]. The film withstood sweeps to ±60 V (a nominal field of 26 MV/cm) without breakdown, confirming that the field in the film is far below the nominal value. PFM is therefore used here to establish local switching and the 180° phase reversal, whereas coercive fields are quantified from the ferroelectric capacitors described below. Additional PFM hysteresis loops and images of a 5 nm thick β-$In_2Se_3$ are shown in Supplementary Figure 14. These results indicate that out-of-plane ferroelectricity is maintained even at a reduced thickness of 5 nm.

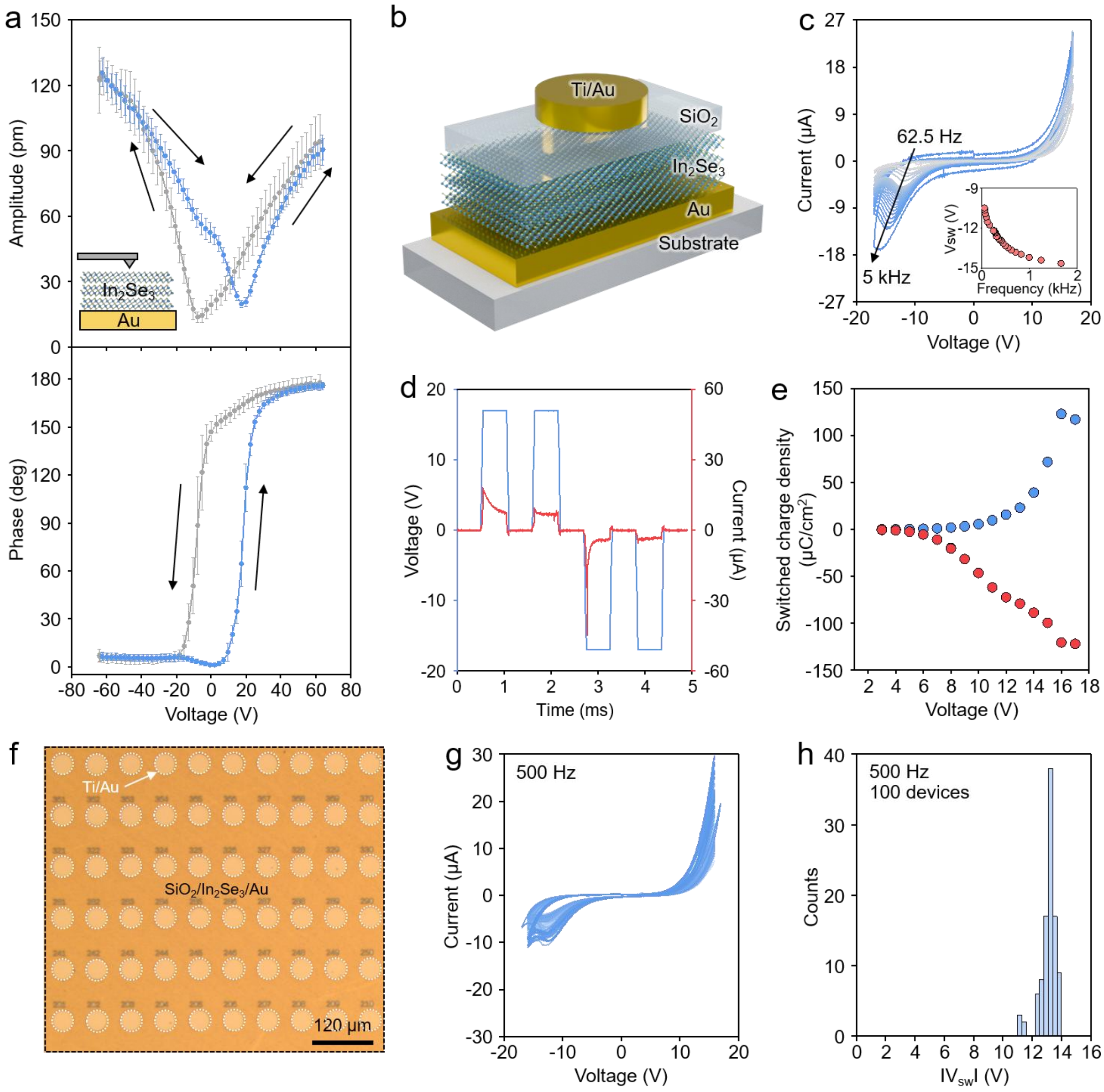

**Figure 2. Emergence of ferroelectricity in β-$In_2Se_3$.** (a) Piezoresponse amplitude and phase as a function of the applied tip bias for a 22.7-nm-thick β-$In_2Se_3$ film, showing characteristic butterfly-shaped amplitude and ferroelectric phase-hysteresis loops. (b) Schematic illustration of the Ti/Au/$SiO_2$ (29.7 nm)/β-$In_2Se_3$ (8.1 nm)/Au capacitor structure. (c) Current–voltage (I–V) hysteresis loops measured at different voltage frequencies. The inset shows the frequency dependence of the switching voltage. (d) Time-resolved current response of the device shown in (b) under applied voltage pulses. (e) Switched charge density extracted from the PUND measurements as a function of pulse amplitude. (f) Optical microscope (OM) image of capacitor array. (g) I–V hysteresis loops measured at 500 Hz from an array of 100 devices. (h) Statistical distribution of the switching voltages (negative switching peak) extracted from the devices shown in (g).

To further evaluate the electrical switching behavior, capacitor devices were fabricated as shown in Figure 2b. The device consists of a Ti/Au top electrode, a 29.7 nm-thick $SiO_2$ layer, an 8.1 nm-thick β-$In_2Se_3$ film and an Au bottom electrode, with the layer thicknesses determined by AFM (Supplementary Figures 15 and 16). The $SiO_2$ layer was introduced to limit leakage through the ultrathin β-$In_2Se_3$ and to prevent premature dielectric breakdown, and the same stack was used for all capacitor measurements in this work. The capacitors are circular with a diameter of 40 μm (Figure 2f). During the measurements, the voltage was applied to the bottom electrode while the top electrode was grounded. Figure 2c shows current–voltage (I–V) hysteresis loops measured using bipolar triangular waveforms from 62.5 Hz to 5 kHz. Pronounced current peaks are observed during the voltage sweep, consistent with ferroelectric polarization switching[29]. Because the applied voltage is divided between β-$In_2Se_3$ and $SiO_2$, the peak voltage is defined as the switching voltage ($V_{SW}$). The magnitude of the negative $V_{SW}$ increases with increasing frequency, from ~10.5 V at 62.5 Hz to ~14.7 V at 1.67 kHz (inset of Figure 2c). This frequency dependence is characteristic of time-dependent ferroelectric switching kinetics[29]. Charge injection and ionic transport can also produce frequency-dependent currents, so the assignment to polarization switching rests on the PUND measurements described next. A reference capacitor without β-$In_2Se_3$, made with the same $SiO_2$, shows a dielectric constant of 3.9 and no capacitance peaks (Supplementary Figure 17). The DC I–V characteristics of the β-$In_2Se_3$ device are shown in Supplementary Figure 18. In addition, the I–V characteristics show a pronounced asymmetry between the two voltage polarities: the switching peak is resolved on the negative branch, whereas on the positive branch it is masked by the larger leakage current of the asymmetric Au/β-$In_2Se_3$/$SiO_2$ stack. A similar asymmetry is observed in the local PFM response in Figure 2a. The dissimilar β-$In_2Se_3$/$SiO_2$ and β-$In_2Se_3$/Au interfaces may further contribute to this asymmetric switching behavior[30].

To quantify the out-of-plane switchable polarization, Positive-Up-Negative-Down (PUND) measurements were performed using the capacitor structure shown in Figure 2b. The time-resolved current response (Figure 2d) exhibits distinct transients during the switching pulses (P and N), while the corresponding non-switching pulses (U and D) show markedly reduced responses, enabling the ferroelectric switching contribution to be isolated. Notably, a clear switching response is also resolved under positive bias, where the feature is less evident in the I–V measurements of Figure 2c. The switched charge was extracted by subtracting the non-switching current and integrating the resulting transient over time[29]. Figure 2e shows the switched charge density extracted from the PUND measurements as a function of pulse amplitude (the P and U transients are overlaid in Supplementary Figure 21 so that the subtraction can be judged). Switching emerges near 8 V and the switched charge increases with pulse amplitude up to ~16 V. Above ~14 V (~1.0 MV/cm in the β-$In_2Se_3$ layer), the switched charge rises steeply and is likely overestimated by leakage and other non-ferroelectric contributions, such as pulse-dependent charge transport and voltage redistribution in the semiconducting ferroelectric/dielectric stack[29]; in a semiconducting ferroelectric, the switched charge also includes the redistribution of screening carriers. Throughout this work we therefore quote, as a conservative value, the positive-branch (smaller) $2P_r$ measured at a field of ~1.0 MV/cm in the $In_2Se_3$ layer (series-capacitor estimate, see below), the lowest field at which the switching peak is fully developed (Figure 2e). We refer to this value as the apparent $P_r$, whereas the larger values at higher amplitude are reported as switched charge densities and not as $P_r$. Because the $SiO_2$ layer is thick and has a low permittivity, the polarization charge must also be compensated for the state to be stable. Without compensation, the calculated out-of-plane polarization of 1.87 μC/cm$^2$ would produce a depolarization field of ~1.2 MV/cm in the β-$In_2Se_3$ layer (series-capacitor estimate), comparable to the coercive field. In the semiconducting β-$In_2Se_3$, this field can be screened by mobile carriers and by charge trapped at the β-$In_2Se_3$/$SiO_2$ interface, and the screening charge also contributes to the PUND charge[8]. The FeS-FETs in Figure 5c,d show that the film can supply this compensation ($1.2 \times 10^{13}$ charges cm$^{-2}$), as their channel charge changes by (1–2) × $10^{13}$ cm$^{-2}$ between the on and off states. An apparent $P_r$ of 11 μC/cm$^2$, by contrast, would need ~7 × $10^{13}$ cm$^{-2}$ and would otherwise produce a depolarization field of ~7 MV/cm, so most of the PUND charge is attributed to screening carriers rather than to lattice polarization.

Figure 2f,g show I–V measurements at 500 Hz on 100 randomly selected capacitors distributed over an area of 4.0 × 1.5 mm$^2$, used to assess device-to-device uniformity. All devices exhibited clear ferroelectric switching with closely matched $V_{SW}$ values, leakage-current levels and I–V loop shapes. The $V_{SW}$ distribution extracted from the negative switching peak (Figure 2h) yields a mean of 13.13 V, a standard deviation of 0.57 V and a coefficient of variation of ~4.3%, demonstrating uniform switching across the array.

Given its importance for device miniaturization and 3D integration, out-of-plane ferroelectricity was further examined at reduced thickness. By decreasing the number of MOCVD growth cycles, the thickness of β-$In_2Se_3$ was scaled down to the trilayer (~2.7 nm) and bilayer (~1.9 nm), and capacitors were fabricated using the same device architecture as those used for the out-of-plane polarization measurements in Figure 2. Notably, clear ferroelectric switching signals persisted down to the bilayer limit (Supplementary Figures 19 and 20). As in Figure 2, the switching peak is resolved on the negative branch, and the 29.7-nm $SiO_2$ layer, which also serves as a leakage barrier for the ultrathin films, carries most of the applied voltage. To compare the switching voltage across different thicknesses, AC–IV characteristics measured at the same pulse frequency are shown in Supplementary Figures 19e and 20e. Because the applied voltage is distributed across both the $SiO_2$ and $In_2Se_3$ layers, the voltage corresponding to each switching peak does not directly represent the voltage across the ferroelectric layer. We therefore extracted the voltage drop across $In_2Se_3$ using the dielectric constant of 29.7 nm-thick $SiO_2$ (k = 3.9, measured on the reference capacitor in Supplementary Figure 17)[31] and $In_2Se_3$ (k = 17)[32] layers, together with the measured $In_2Se_3$ thickness of each device. The coercive field was subsequently calculated by dividing the extracted voltage by the corresponding film thickness. The extracted values (0.91, 0.98 and 0.97 MV/cm for the 1.9, 2.7 and 8.1 nm films) remain nearly constant over the investigated thickness range, which, within the series-capacitor model, corresponds to a coercive field that does not increase upon scaling (Figure 3a). These values rest on the dielectric constant of bulk $In_2Se_3$, because that of β-$In_2Se_3$ has not been measured. Because most of the applied voltage drops across the $SiO_2$ layer, they are nearly inversely proportional to the assumed k (they would be ~1.6 MV/cm for k = 9.7), and the smaller dielectric constants reported for flakes thinner than about six quintuple layers[32] would raise the values of the bilayer and trilayer accordingly.

Figure 3a compares these values with the coercive fields reported for other ferroelectric materials[33-43]. In addition to the coercive field, we evaluated the thickness-dependent $P_r$ using PUND measurements. For devices of each thickness, the switched charge increased with the applied voltage (Figure 2e and Supplementary Figures 19f and 20f). To minimize overestimation arising from leakage, we used the conservative positive $P_r$ defined above (measured at ~1.0 MV/cm, just above the coercive field) for this comparison. The corresponding $P_r$ values were benchmarked against those of conventional 3D ferroelectrics[30,36-38,41,44-51] and other 2D ferroelectric materials[34,52-54] (Figure 3b), where only measured values are compared. Two features stand out. Below 3 nm, where vertical switching with a quantified coercive field has so far been reported mainly in sliding ferroelectrics (bilayer hBN and 3R-$MoS_2$)[33,34], 2D $Ga_2O_3$[43] and ultrathin HZO, β-$In_2Se_3$ shows the lowest coercive field (0.9–1.0 MV/cm, series-capacitor estimate with k = 17, versus ~2–14 MV/cm), and its apparent $P_r$, an upper bound on the lattice polarization (1–2 $\mu C/cm^2$, see above), is of the same order as that of HZO at a similar thickness[37,40]. Furthermore, whereas the coercive fields of HZO (2–5 MV/cm at ≤5 nm)[37,40] and AlScN (above 2 MV/cm at any thickness)[29,48-51] remain higher, the extracted coercive field of β-$In_2Se_3$ is independent of thickness down to the bilayer. In a stack without a series dielectric, the switching voltage would therefore scale with thickness, as required for low-voltage scaled devices. Together with the preserved coercive field, this thickness-resilient response makes β-$In_2Se_3$ attractive for scaled, low-voltage ferroelectric devices.

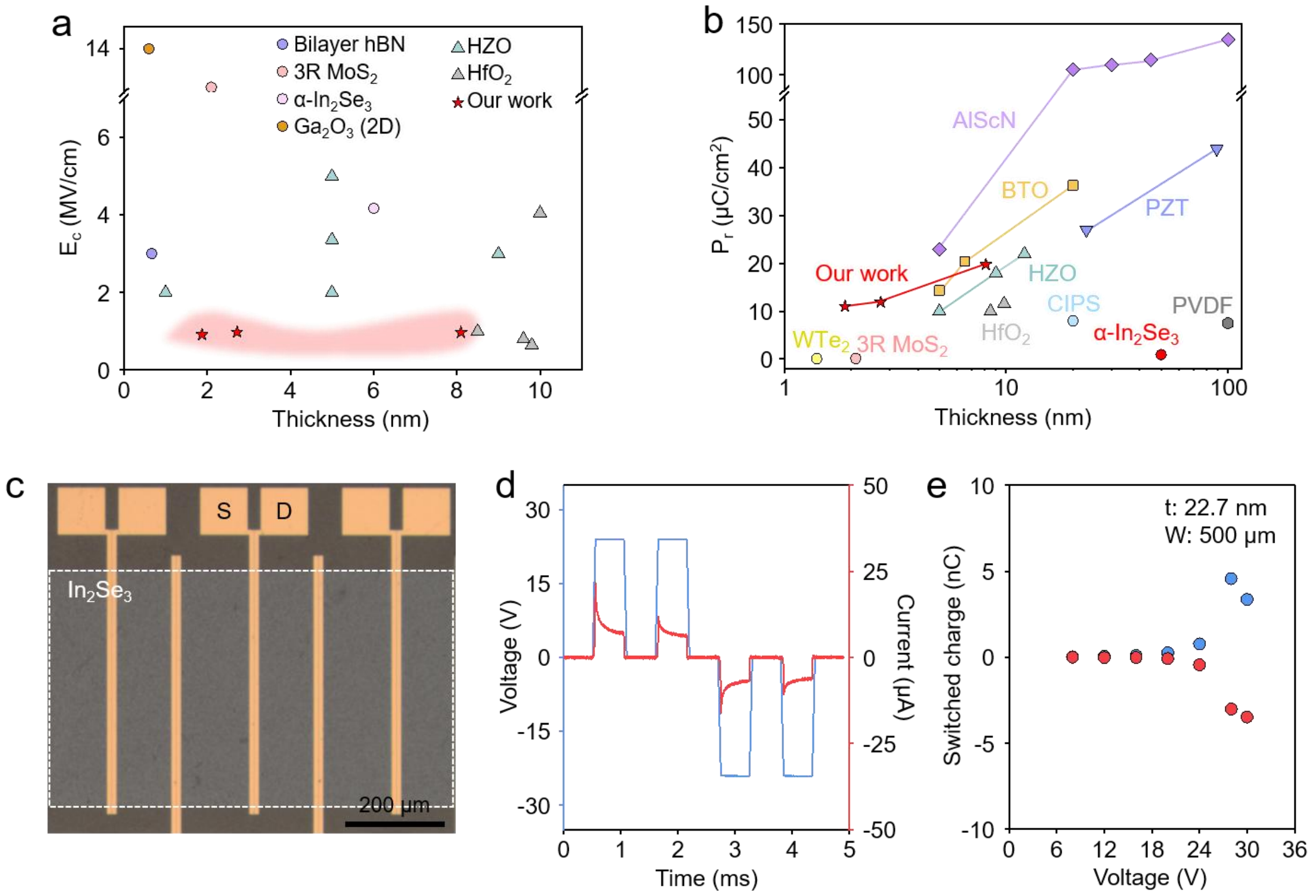

**Figure 3. Thickness scaling of out-of-plane ferroelectricity and in-plane switching in β-$In_2Se_3$.** (a) Benchmark of coercive field as a function of ferroelectric-layer thickness for $In_2Se_3$ and representative ferroelectric materials. (b) Benchmark of the measured remanent polarization as a function of ferroelectric-layer thickness for $In_2Se_3$ and representative ferroelectric materials. All values in (a) and (b) are experimental; calculated polarizations are not included. (c) OM image of the lateral device fabricated on a 22.7-nm-thick $In_2Se_3$ film (electrode gap, 500 nm; channel width, 500 μm). (d) Time-resolved current response of the device shown in (c) under applied voltage pulses. (e) Switched charge obtained from the PUND sequence in (d) as a function of pulse amplitude. The charge is dominated by conduction in the channel and is not converted to a polarization (see text).

In-plane switching was also examined using lateral devices in which two Ti/Au electrodes separated by a 500-nm gap were patterned on a 22.7-nm-thick β-$In_2Se_3$ channel (width, 500 μm) on insulating sapphire, so that the field in the gap lies in the plane of the film (Figure 3c). The resulting AC–IV and DC–IV characteristics showed well-defined hysteretic behavior (Supplementary Figure 22). Consistent with this response, PUND measurements displayed distinct current transients during the P and N switching pulses (Figure 3d). As the pulse amplitude was increased stepwise, the charge remaining after subtraction of the non-switching pulses stayed below ~1 nC up to 24 V and rose to 3–5 nC at 28–30 V for both polarities, which corresponds to a threshold at a nominal lateral field of ~0.5 MV/cm (Figure 3e). This charge is about three orders of magnitude larger than the ~5 pC expected for the calculated in-plane polarization of 21 μC/cm$^2$ and the cross-section of the channel. It is therefore dominated by a conduction current that differs between the switching and non-switching pulses, as expected when the conductance of the semiconducting channel depends on its polarization state (Supplementary Figure 22), and we report it as a switched charge without converting it to a remanent polarization. Lateral hysteresis in the β′ polymorph has been attributed to the field-driven migration of Se vacancies[55]. In the present devices the switched charge appears only above a threshold field and is reversed within the sub-millisecond pulses of both polarities (Figure 3d, e), and the in-plane displacement resolved by STEM is a uniform distortion of the In–Se framework rather than a vacancy-ordered superstructure (Supplementary Figures 7 and 8). These observations are consistent with switching of the in-plane polarization component and argue against slow ionic drift as the origin of the lateral hysteresis, but they do not yield a quantitative in-plane polarization.

After confirming out-of-plane ferroelectric switching and its thickness scaling in β-$In_2Se_3$, we further evaluated whether the polarization reversal could be sustained during repeated electrical cycling. Following the endurance protocol established in our previous work[29], two cycling conditions were examined. Figure 4a shows the endurance test in which the applied voltage was adjusted during cycling to maintain an apparent $2P_r$ of approximately 10 μC/cm$^2$. The applied voltage started at 12 V, rose to ~13.4 V within the first ~50 cycles and decreased to 10 V at $10^7$ cycles (lower panel of Figure 4a). The gray scatters represent the apparent $2P_r$ (the switching charge minus the non-switching charge) and the blue scatters the upper-bound $2P_r$, which is defined from the total charge of the switching pulses without subtraction[29]. Both polarization metrics show a similar overall evolution over the cycling range. The apparent $2P_r$ remained near the target value up to ~$10^7$ cycles, indicating that the out-of-plane polarization reversal persists during repeated switching.

Figure 4b further examines the cycling behavior under a fixed applied voltage of 12 V. Unlike the controlled polarization condition in Figure 4a, $2P_r$ evolves with cycle number under the fixed voltage. The polarization response initially remains relatively stable and then decreases over ~$10^3$-$10^5$ cycles, consistent with the polarization fatigue commonly observed in conventional ferroelectric capacitors[29,56-58]. At higher cycle numbers the apparent $2P_r$ increases again; because the upper-bound $2P_r$, which includes the non-switching current, rises in parallel, this increase can be attributed to rising leakage as the device approaches breakdown rather than to a recovery of the polarization. A clear polarization response nevertheless persists up to ~$6 \times 10^5$ cycles at a fixed 12 V. To further investigate the increase in $2P_r$ near the end of cycling, as shown in Figure 4b, we performed an additional endurance measurement at a reduced voltage of 8 V, as shown in Figure 4c. This voltage lies at the onset of switching and well below $V_{SW}$ ≈ 13 V, so that only part of the polarization (3–4 μC/cm$^2$, about one third of the 11–13 μC/cm$^2$ switched at 12 V in Figure 4b) is reversed in each cycle; Figure 4c therefore characterizes sub-coercive partial switching rather than full-switching endurance. The apparent $2P_r$ remained nearly constant at approximately 3-4 μC/cm$^2$ for up to 2.1 × 10$^7$ cycles, and no increase in apparent 2Pr was observed near the end of cycling. These results suggest that the increase in apparent $2P_r$ at higher voltage originates from enhanced non-switching current contributions under higher electrical stress. In contrast, the lower voltage enables stable partial switching with minimal variation in $2P_r$ over 2.1 × 10$^7$ cycles.

Figure 4d evaluates the retention of the switched polarization in β-$In_2Se_3$ using a test voltage of 14 V. The pulse configuration used for the retention measurement is shown in the inset. Two reset pulses were first applied to initialize the polarization state, followed by two set pulses to switch the polarization. After a programmed delay, two read/check pulses were applied to evaluate the retained polarization. The delay time was systematically varied to examine the polarization response as a function of retention time. No loss of polarization is observed within the measured range, which extends to 10$^4$ s for the positive state and to 1.3 × 10$^5$ s (36 h) for the negative state: the positive-branch $2P_r$ stays at 7–8 μC/cm$^2$, and the magnitude of the negative-branch $2P_r$ does not decrease (it drifts from ~19 to ~25 μC/cm$^2$). Furthermore, extrapolation of the retained polarization projects a retention of over ten years. Dangling-bond-free van der Waals surfaces may reduce interfacial charge trapping and associated depolarization-field-induced polarization loss[6,59]. Figure 4e shows representative current responses recorded during the retention measurement. In both the reset and set sequences, pronounced differences between the first and second pulse responses indicate polarization switching. In contrast, the two test pulses applied after the retention interval exhibit nearly identical current responses, indicating that no additional polarization switching occurs and supporting the retention of the programmed polarization state observed in Figure 4d.

Figure 4f compares the endurance and retention characteristics of β-$In_2Se_3$ with those of representative ferroelectric materials[9,29,33,34,44,46,52,53,60-65] (measured values; retention at the ten-year line is extrapolated for all materials; see also Supplementary Table 4). Among 2D ferroelectrics, exfoliated sliding-ferroelectric stacks reach an endurance beyond 10$^{11}$ cycles but with retention measured only to ~10$^6$ s[33,34], and the 2D ferroelectrics projected to retain for ten years have been cycled no more than ~10$^4$ times. β-$In_2Se_3$ combines an endurance of 10$^7$ cycles with a projected ten-year retention in a wafer-scale film with device-to-device statistics (Figure 2f–h), a combination otherwise available only in HZO and AlScN[29,44], neither of which is a semiconductor. These results indicate stable non-volatile polarization in β-$In_2Se_3$ under both prolonged storage and repeated electrical switching.

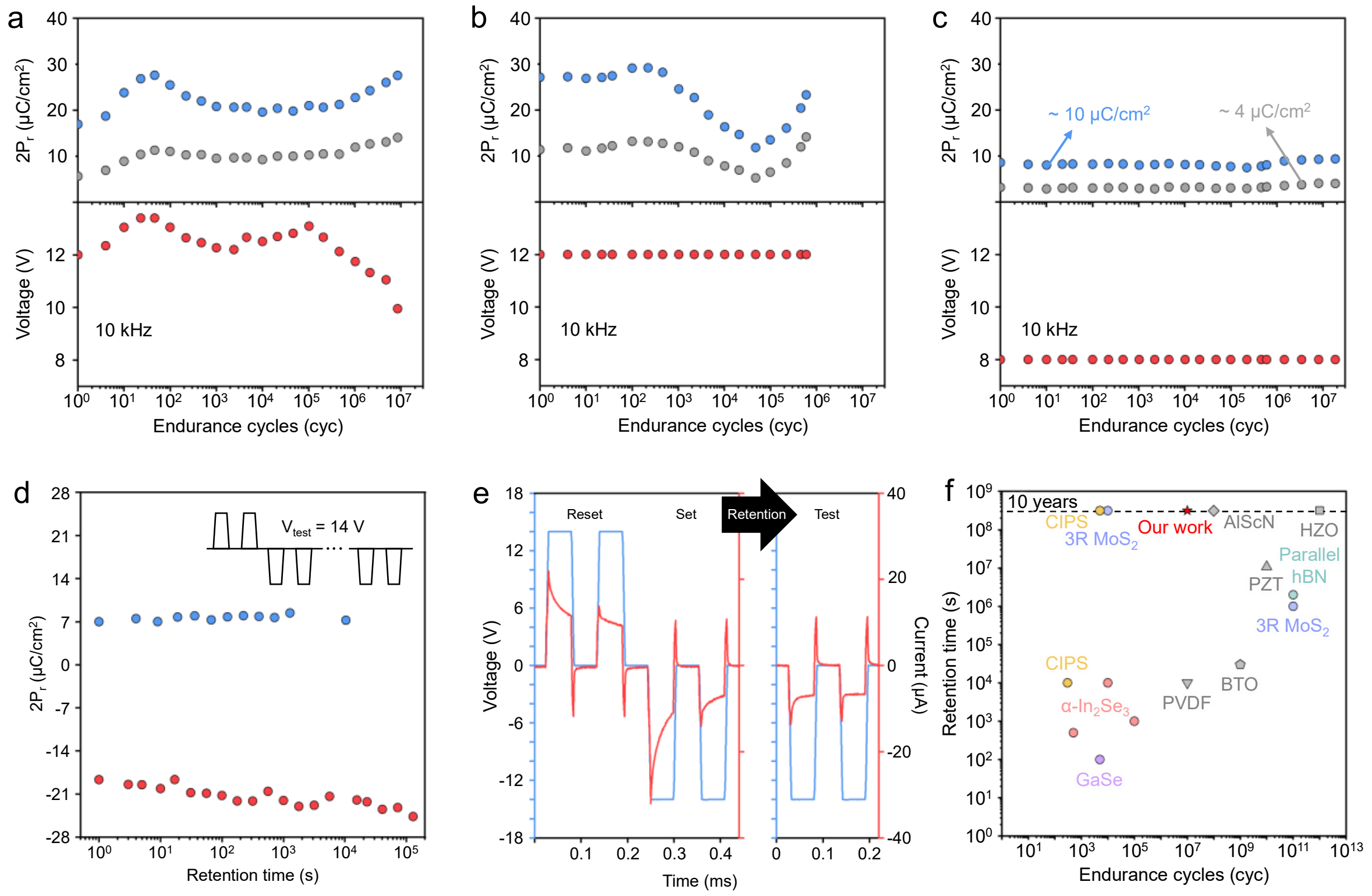

**Figure 4. Nonvolatile memory characteristics of β-$In_2Se_3$.** (a) Endurance of the capacitor device shown in Figure 2 under adjusted applied voltage pulses to maintain a constant apparent $2P_r$. (b, c) Endurance under constant applied voltage pulses of 12 and 8 V, respectively. (d) Retention at a test voltage of 14 V, measured up to $10^4$ s for the positive state and $1.3 \times 10^5$ s for the negative state. (e) Representative voltage and current responses during the retention measurement, showing the reset and set pulses and the test pulses after a retention time of $1.56 \times 10^4$ s. The sharp current spikes at the pulse edges are capacitive charging currents; they are the same for the switching and non-switching pulses and cancel in the subtraction. (f) Benchmark of non-volatile memory characteristics, including endurance and retention, across representative ferroelectric materials. Retention at the ten-year line is extrapolated for all materials, and the endurance of β-$In_2Se_3$ is taken from (a).

Finally, we fabricated top-gated ferroelectric semiconductor field-effect transistors (FeS-FETs) using $In_2Se_3$ as the active channel (Figure 5a), following the device concept introduced for mechanically exfoliated α-phase $In_2Se_3$[8]. Unlike conventional FeFETs, in which the ferroelectric layer serves as the gate dielectric, the ferroelectric semiconductor itself functions as the channel in FeS-FETs. The dangling-bond-free van der Waals surface of $In_2Se_3$ minimizes interfacial oxide formation and depolarization effects, which is expected to stabilize the polarization states (Figure 5b). $HfO_2$ (k ≈ 15-20) and $ZrO_2$ (k ≈ 20-25), both 25 nm thick and deposited on a thin $SiO_x$ seed layer (Methods), were employed as gate dielectrics on nominally identical $In_2Se_3$ channels (channel thickness: 4.5 nm); the two stacks have similar equivalent oxide thicknesses (EOT ≈ 4.9–6.5 nm for $HfO_2$ and 3.9–4.9 nm for $ZrO_2$, excluding the seed layer) (see Supplementary Figure 23 for details of the fabrication process).

The $HfO_2$-gated device exhibits clockwise hysteresis (Figure 5c), whereas the $ZrO_2$-gated device shows counterclockwise hysteresis (Figure 5d), and in both devices the memory window widens with the gate-voltage range without a change in the sense of the loop (Supplementary Figure 24 and Supplementary Tables 1 and 2). For an n-type channel, electron trapping at the gate interface produces clockwise hysteresis, so the counterclockwise loop of the $ZrO_2$-gated device cannot be explained by channel trapping and is the response expected for a ferroelectric semiconductor channel, in which the polarization-bound charge sets the carrier density of the surface that carries the current (Figure 5b,f)[8]. Because the two gate stacks have similar EOTs and the sense of the loop does not depend on the sweep range, the difference between the two devices is set by the gate interface rather than by the field in the channel. For the $HfO_2$-gated device, the clockwise loop can arise either from polarization-induced modulation of the gate-side surface, which is depleted when the polarization points away from the gate (Figure 5e), or from electron trapping at the $HfO_2/SiO_x$ interface, and the present data do not separate the two contributions. The $HfO_2$-gated device, whose stack sustained the ±8 V program and erase pulses without breakdown, was used to evaluate the memory characteristics of the $In_2Se_3$ FeS-FET. The two states remain distinguishable after $10^4$ s, with an on/off ratio above $10^2$ (Figure 5g), and over $10^4$ program/erase cycles, with a ratio of ~4 (Figure 5h), which demonstrates non-volatile two-state operation in a semiconducting ferroelectric channel. The device also responded to much shorter pulses than the 900 ms used in Figure 5h: erase pulses of +8 V as short as 200 ns reduced $I_{ds}$ by ~24%, program pulses of −20 V lasting 100 μs to 1 ms increased $I_{ds}$ by 20–60%, and the change grew with pulse width. With ±10 V, 90-ms pulses, the two states remained distinguishable for more than $10^4$ cycles (Supplementary Figure 25). Because one wafer-scale film serves as both the switching layer of a capacitor and the channel of a transistor, β-$In_2Se_3$ offers a single-material route to the ferroelectric semiconductor devices envisaged for in-memory and in-sensor computing[8-10].

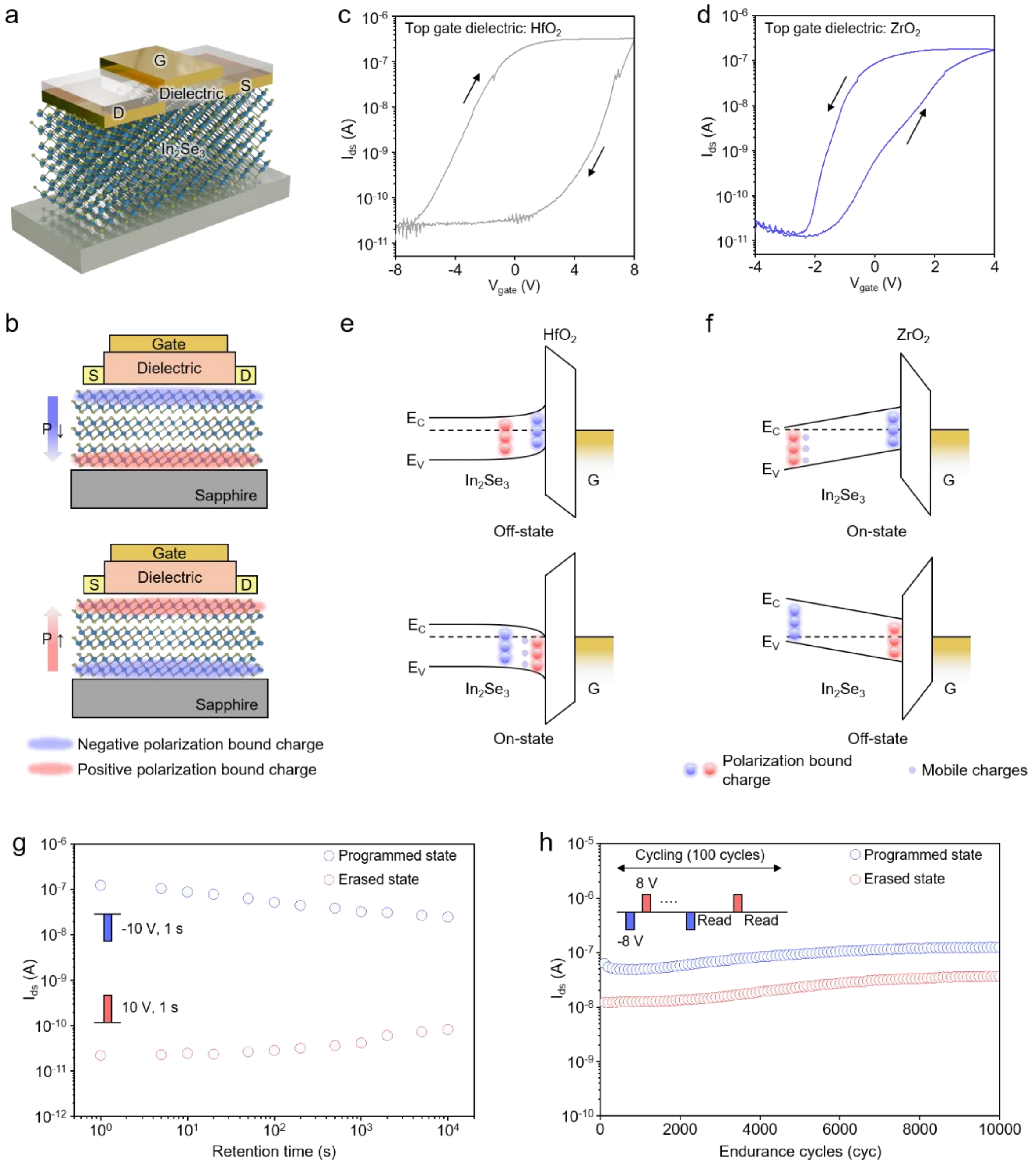

**Figure 5. Demonstration of β-$In_2Se_3$ ferroelectric semiconductor field-effect transistors (FeS-FET).** (a) Schematic of the β-$In_2Se_3$ FeS-FET with top gate structure. (b) Schematic illustration of the polarization-bound charge distribution in the FeS-FET under polarization down (after applying a positive gate bias) and polarization up (after applying a negative gate bias) states. (c) Transfer characteristics of the device employing $HfO_2$ as the gate dielectric, measured at $V_{ds}$ = 1 V. (d) Transfer characteristics of the device employing $ZrO_2$ as the gate dielectric, measured at $V_{ds}$ = 1 V. (e) Band diagrams of the $HfO_2$-gated device, illustrating the on and off states when transport is dominated by the gate-side surface of the channel. (f) Band diagrams of the $ZrO_2$-gated device, illustrating the on and off states when transport is dominated by the lower part of the channel. (g) Retention characteristics of the device employing $HfO_2$ as the gate dielectric. The read operation was performed at $V_{ds}$ = 1 V, $V_{gate}$ = 0 V. (h) Endurance characteristics of the same device shown in (g). Program and erase operations were performed using -8 V and +8 V with a duration of 900 ms, respectively, while the read operation was conducted at $V_{ds}$ = 1 V, $V_{gate}$ = 0 V.

## Conclusion

β-$In_2Se_3$, the high-temperature polymorph of indium selenide that has long been classified as paraelectric, hosts a bistable polar state defined by picometre-scale displacements within each quintuple layer. The state is switched by a vertical field, survives down to the bilayer (1.9 nm) with an unchanged coercive field of 0.9–1.0 MV/cm, and, in wafer-scale films, supports capacitor arrays with a switching-voltage spread below 5%, an endurance of $10^7$ cycles and transistors in which the ferroelectric itself is the channel. The switched polarization and low coercive field retained at this thickness distinguish β-$In_2Se_3$ from many ferroelectric systems that deteriorate rapidly upon thinning. Because the polar state coexists with a semiconducting gap, a single wafer-scale film provides both the storage layer of a capacitor and the channel of a transistor. This combination of wafer-scale growth, quantified coercive field and polarization, device-to-device statistics and a ferroelectric-channel transistor has not, so far, been reported for any other 2D ferroelectric. More broadly, these results suggest that nominally nonpolar layered compounds may harbor electrically accessible structural degrees of freedom that have remained unexplored. Harnessing such hidden instabilities could expand the materials space available for energy-efficient memory and integrated electronic technologies.

## Methods

*MOCVD growth of $In_2Se_3$:*

β-$In_2Se_3$ films were synthesized in a showerhead-type vertical metal–organic chemical vapor deposition (MOCVD) system equipped with an inductively heated TaC-coated graphite susceptor (SMI, Inc.). Trimethylindium (TMIn; $(CH_3)_3In$) and dimethyl selenide (DMSe; $(CH_3)_2Se$) were used as the In and Se precursors, respectively, with $H_2$ serving as the carrier gas. The two precursor streams were supplied independently and mixed only after entering the reaction chamber to minimize parasitic gas-phase reactions. Precursor delivery was regulated using computer-controlled pneumatic valves, mass-flow controllers, pressure controllers and bubbler-temperature controllers, allowing precise control over the precursor flux and pulse sequence. The TMIn and DMSe bubblers were maintained at 32 °C, corresponding to vapor pressures of approximately 3 and 326 Torr, respectively[66,67]. Growth was carried out at a temperature of 525 °C and a total reactor pressure of 100 Torr. During deposition, the substrate was rotated at approximately 50 rpm on the TaC-coated graphite susceptor, while a constant $H_2$ flow of 50 sccm was maintained. To obtain β-$In_2Se_3$, a phase-selective growth strategy was employed following our previously established MOCVD protocol[15]. The growth phase was controlled by adjusting both the maximum Se/In precursor ratio and the relative DMSe delivery duration within each precursor cycle. These parameters were tuned to access the β-$In_2Se_3$ growth window shown in Supplementary Figure 1. In each growth cycle, the TMIn and DMSe delivery durations were $t_i$ = 45 s and $t_d$ = 60 s, respectively ($t_d/(t_i + t_d)$ = 0.57; Supplementary Figure 1b,c), and $H_2$ flows of 9 and 3 sccm through the TMIn and DMSe bubblers gave a maximum Se/In ratio of ~24. The 1.9, 2.7, 4.5, 8.1 and 22.7 nm films were grown with 42, 66, 72, 81 and 165 cycles, respectively. The thickness did not scale linearly with the number of cycles, and the thickness of each film was therefore determined by AFM.

*Device fabrication:*

For PFM measurements and vertical capacitor fabrication, a 100-nm-thick Au layer was deposited onto the as-grown $In_2Se_3$ film on sapphire using an electron-beam evaporator (PVD 75 electron-beam evaporator, Kurt J. Lesker Company). An epoxy layer was subsequently coated on the Au surface, followed by placement of a Si substrate on top. After curing the epoxy, the sapphire growth substrate was mechanically detached, and the resulting stack was flipped to obtain an $In_2Se_3$ film supported by Au. This transferred structure was used for PFM characterization. Vertical capacitors were fabricated on the Au-supported $In_2Se_3$ films. The 29.7-nm $SiO_2$ layer was deposited at room temperature in the same electron-beam evaporator. Top electrodes (circular, 40 μm in diameter) were defined by electron-beam lithography (EBPG5200 +, Raith), followed by sequential deposition of Ti (40 nm) and Au (80 nm) to form the metal contacts, as illustrated in Supplementary Figure 12. Lateral capacitors were fabricated directly from $In_2Se_3$ films grown on sapphire. The $In_2Se_3$ layer was first patterned into rectangular channels by electron-beam lithography and subsequently etched by reactive-ion etching. After removal of the electron-beam resist, a second electron-beam lithography step was used to define the lateral electrodes (two electrodes, each 10 μm long in the current direction and 500 μm wide, separated by a 500-nm gap), followed by sequential Ti/Au (40/80 nm) deposition to complete the device fabrication. For FeS-FET fabrication, $In_2Se_3$ films grown on sapphire were patterned into rectangular channels by electron-beam lithography and reactive-ion etching. The source and drain electrodes were defined, followed by deposition of Ti/Au (20/20 nm) by electron-beam evaporation. For the top-gate stack, a thin Si layer was deposited by electron-beam evaporation and naturally oxidized, followed by atomic layer deposition (S200 system, Cambridge Nanotech) of the gate dielectric. Finally, Cr/Au (10/30 nm) top-gate electrodes were deposited to complete the top-gated FeS-FETs (Supplementary Figure 23).

*Characterization:*

Raman spectra were acquired using a confocal Raman microscope (LabRAM HR Evolution, HORIBA) equipped with a 600 grooves $mm^{-1}$ grating and a 633 nm continuous-wave laser as the excitation source. The Raman signal was collected using a CCD detector. The surface morphology of the $In_2Se_3$ films was characterized by atomic force microscopy (AFM; Dimension Icon, Bruker) operating in tapping mode. X-ray photoelectron spectroscopy (XPS) was performed on a K-Alpha spectrometer (Thermo Fisher Scientific) using monochromatic Al Kα radiation (1,486.6 eV) under ultrahigh vacuum (~$10^{-9}$ Torr), and binding energies were calibrated to the C 1s peak at 284.8 eV. Cross-sectional transmission electron microscopy (TEM) lamellae were fabricated by in situ lift-out in a $Ga^+$ focused ion beam (FIB)/scanning electron microscopy (SEM) dual-beam instrument (Helios 5 UX, Thermo Fisher Scientific) after depositing Pt and C protective layers. PFM measurements were performed using a Vero AFM (Oxford Instruments Asylum Research) with a Pt/Ir-coated conductive Si probe (PPP-EFM, NANOSENSORS). A fixed excitation frequency of 347.28 kHz, the measured contact-resonance frequency, was used. A write bias was applied through the conductive tip, and the off-field piezoresponse was obtained during the subsequent zero-bias read intervals. Electrical characterization, including DC and AC current–voltage measurements, PUND measurements, retention and endurance tests, was performed at room temperature using a probe station interfaced with a Keithley 4200A-SCS semiconductor parameter analyzer (Tektronix, Inc.). Capacitance-voltage measurements were performed using the capacitance-voltage unit (CVU) of the Keithley 4200A-SCS in the parallel capacitance-dissipation factor (Cp–D) mode. The capacitance was recorded during forward and reverse DC voltage sweeps using a 30 mV AC excitation at 50 kHz. Capacitance–frequency characteristics of the reference capacitor were recorded from 10 kHz to 1 MHz.

*HAADF-STEM imaging:*

HAADF-STEM images were acquired with a double $C_s$-corrected STEM (JEM-ARM200CF, JEOL) operated at 200 kV with a probe convergence semi-angle of 24 mrad, a HAADF detector collection angle range of 68–280 mrad (camera length 80 mm), and a probe current of 23 pA.

The specimens were aligned to the $[11\bar{2}0]$ zone axis of β-$In_2Se_3$ with the c axis parallel to the vertical image axis. To minimize distortion from specimen drift, each image was recorded as a series of four frames of 1024 × 1024 pixels (dwell time, 2 μs per pixel; 0.09917 Å per pixel from the microscope calibration recorded in the image metadata) that were rigidly registered and averaged using SmartAlign (HREM Research Inc.)[68]. The registered images were cropped to the field of view common to all frames (965–1020 × 980–1011 pixels), from which 870 × 870-pixel regions (≈8.6 × 8.6 $nm^2$) were used for the analysis. Two series were acquired for each state with the fast-scan direction rotated by 90°, so that the out-of-plane and in-plane atomic distances could each be measured from an image whose fast-scan direction was parallel to the measured component (Supplementary Figures 5-8). This scheme minimizes the influence of scan distortion, which arises predominantly along the slow-scan direction[69,70]. Statistical noise in the averaged images was reduced with a local two-dimensional Wiener filter (HREM-Filters Pro, HREM Research Inc.) in DigitalMicrograph (Gatan) using identical settings for the Program and Erase images.

*qSTEM simulation:*

HAADF-STEM images were simulated with the multislice code qSTEM (v2.50)[71] to provide a reference for the atomic-distance measurements. The input model was the DFT-relaxed centrosymmetric structure of β-$In_2Se_3$ (space group $R\bar{3}m$; see DFT calculations), oriented with the electron beam along $[11\bar{2}0]$ and expanded to a 12 × 12 × 5 supercell (≈12 × 12 $nm^2$, 5.0 nm thick). The electron-optical parameters were matched to the experimental conditions (200 kV; convergence semi-angle 24 mrad; defocus 0 nm; $C_3 = -0.037$ mm; $C_c = 1.0$ mm; $\Delta E = 0.6$ eV; HAADF detector 68–280 mrad). The probe was sampled on a 400 × 400 array with a sampling interval of 0.05 Å, and the specimen was divided into 25 slices, each 2.0 Å thick. The central 100 × 100 $Å^2$ region was scanned with 256 × 256 pixels. The simulated image was interpolated fourfold to 1024 × 1024 pixels (0.09766 Å per pixel), rotated to the experimental orientation, and cropped to 850 × 850 pixels (≈8.3 × 8.3 $nm^2$). The resulting image (Supplementary Figure 4) was analyzed with the same procedure as the experimental images.

*Atomic-position extraction and projected atomic distance (PAD) analysis:*

Interatomic distances within each quintuple layer (QL) were quantified as projected atomic distances (PADs) between adjacent atomic columns of the Se1–In1–$Se_C$–In2–Se2 sequence (Se1 and Se2, bottom and top Se; In1 and In2, lower and upper In; $Se_C$, central Se), resolved into the components parallel to the layers (in-plane PAD) and to the c axis (out-of-plane PAD). This approach is analogous to the unit-cell-level lattice-parameter mapping from atomic-column positions previously applied to perovskite oxide films[72,73]. The Wiener-filtered experimental images (see HAADF-STEM imaging) and the unfiltered simulated image were processed identically. Atomic-column positions were extracted with HyperSpy (v2.4.0)[74] and Atomap (v0.4.2)[75]. The In columns were located by peak finding (minimum separation 18 pixels ≈ 1.8 Å, relative threshold 0.3) and refined by center-of-mass and two-dimensional Gaussian fitting. Their fitted intensities were then subtracted from the image, and the Se columns were located and refined in the residual image (minimum separation 9 pixels ≈ 0.9 Å). Sequences were constructed by a KD-tree nearest-neighbor search starting from Se1 and connecting, at each step, the nearest column of the other species within 40 pixels (≈4 Å) and ±25° of the projected bond direction, that is, the direction from Se1 to Se2 within a QL in the $[11\bar{2}0]$ projection, which is fixed by the crystal structure and independent of the polarization state. Only complete five-column sequences with a unique $Se_C$ were retained and assigned to QLs by the c-axis coordinate of $Se_C$. Each PAD component was measured in an image acquired with the fast-scan direction parallel to that component (out-of-plane, Supplementary Figures 5 and 6; in-plane, Supplementary Figures 7 and 8), and distances were converted to ångströms using the pixel sizes given above (0.09917 Å per pixel, experimental; 0.09766 Å per pixel, simulated). PAD maps were generated by assigning each PAD value to the rectangle spanned by the two atomic columns, averaging overlapping pixels, and overlaying the result on the HAADF image in false color. PADs are quoted as the sample mean ± standard deviation. The in-plane and out-of-plane displacements of the central Se from the midpoint between In1 and In2 were calculated for each sequence as half the difference between the In1–$Se_C$ and $Se_C$–In2 PADs of the corresponding component and averaged over all sequences. The displacement defined in this way is analogous to the polar cation displacements quantified from atomic-column positions in ferroelectric and polar oxide films and heterostructures[73,76-78]. Uncertainties are given as the standard error of the mean, and positive values denote a shift of $Se_C$ toward In2 in each component. Absolute in-plane PADs carry a residual scan-magnification uncertainty between images, so that comparisons between states use the difference Δ measured within the same image.

*First-principles calculation:*

First-principles calculations were performed using the Vienna Ab initio Simulation Package

(VASP)[79] with the projector augmented-wave (PAW)[80,81] method. The In 4d electrons were treated as valence states, and the HSE06 hybrid functional[82,83] with the D4 dispersion correction[84,85] was employed throughout with a plane-wave cutoff of 500 eV and a Γ-centered 6×6×2 k-mesh. The lattice parameters were fixed to the STEM-measured values (a = 0.400 nm, c = 2.85 nm), and atomic positions were relaxed without symmetry constraints until the residual forces fell below 0.005 eV/Å. The switching path connecting the two polar states was constructed by linear interpolation of atomic coordinates through the centrosymmetric reference structure ($\lambda = 0$). The band gap remained open along the entire path, allowing the polarization to be tracked continuously using the Berry phase approach[86,87]. Band structures were calculated along the high-symmetry path of the monoclinic (Cm) Brillouin zone[88].

**Data availability**

The data that support the findings of this study are available from the corresponding authors upon reasonable request.

**Code availability**

The code used for atomic-position extraction and projected atomic distance analysis is available from the corresponding authors upon reasonable request.

**Author Contributions**

D.J., S.S., K.Y., H.C. and J.J. conceived the idea and designed the work. K.Y. performed material growth and device fabrication. K.Y. and H.C. performed all the electrical measurements. H.C. performed PFM measurement and analysis. K.Y. and J.J. performed materials characterization. T.H.P., H.J.K., J.-H.L. and C.W.M. performed the first-principles calculations. S.R. and Y.-M.K. performed and analyzed the STEM experiments. K.Y., H.C., J.J., S.S. and D.J. jointly analyzed the data and wrote the paper with input from H.S.S. and R.H.O.

**Acknowledgements**

K.Y., H.C., S.S., R.H.O. and D.J. acknowledge support from the National Science Foundation Future of Semiconductors (FuSe) program (Award 2328743) and from the Defense Advanced Research Projects Agency (DARPA) [TWEED Program HR0011-21-9-0055]. This work was supported by the National Research Foundation (NRF) of Korea (Grant No. RS-2025-00516532, RS-2026-25607606, RS-2026-25622986) funded by the Ministry of Science and ICT, by the Institute for Basic Science (IBS-R036-D1), and by the Korea Basic Science Institute (National Research Facilities and Equipment Center) grant funded by the Korea government (MSIT) (No. RS-2025-00573435). This work was also supported by the Korea Basic Science Institute (National Research Facilities and Equipment Center) grant funded by the Korea government (MOE) (No. RS-2026-25536476). C.W.M. acknowledges the support provided by the National Research Foundation of Korea (NRF) grant funded by the Korean government (MSIT) (RS-2022-NR072058, RS-2023-00257666). The authors are grateful for the computational resources at the Korea Institute of Science and Technology Information (KISTI) (KSC-2025-CRE-0125). Computational work for this research was also partially performed on the Olaf cluster supported by IBS Research Solution Center. Y.-M.K. was supported by the National Research Foundation of Korea (NRF) grant (RS-2023-NR076943) funded by the Korean government in Korea.

**Competing interests**

D.J. and R.H.O. are co-founders of Agni Semiconductor, which develops ferroelectric nitride memory. The other authors declare no competing interests.

**References**

1 Schroeder, U., Park, M. H., Mikolajick, T. & Hwang, C. S. The fundamentals and applications of ferroelectric HfO2. *Nature Reviews Materials* **7**, 653–669 (2022). https://doi.org/10.1038/s41578-022-00431-2

2 Kim, K.-H., Karpov, I., Olsson III, R. H. & Jariwala, D. Wurtzite and fluorite ferroelectric materials for electronic memory. *Nature Nanotechnology* **18**, 422–441 (2023). https://doi.org/10.1038/s41565-023-01361-y

3 Mehta, R., Silverman, B. & Jacobs, J. Depolarization fields in thin ferroelectric films. *Journal of Applied Physics* **44**, 3379–3385 (1973). https://doi.org/10.1063/1.1662770

4 Junquera, J. & Ghosez, P. Critical thickness for ferroelectricity in perovskite ultrathin films. *Nature* **422**, 506–509 (2003). https://doi.org/10.1038/nature01501

5 Stengel, M., Vanderbilt, D. & Spaldin, N. A. Enhancement of ferroelectricity at metal–oxide interfaces. *Nature Materials* **8**, 392–397 (2009). https://doi.org/10.1038/nmat2429

6 Leblanc, C., Song, S. & Jariwala, D. 2D Ferroelectrics and ferroelectrics with 2D: Materials and device prospects. *Current Opinion in Solid State and Materials Science* **32**, 101178 (2024). https://doi.org/10.1016/j.cossms.2024.101178

7 Song, S. *et al.* Indium selenides for next-generation low-power computing devices. *Nature Reviews Electrical Engineering* **3**, 185–201 (2026). https://doi.org/10.1038/s44287-025-00251-w

8 Si, M. *et al.* A ferroelectric semiconductor field-effect transistor. *Nature Electronics* **2**, 580–586 (2019). https://doi.org/10.1038/s41928-019-0338-7

9 Wang, S. *et al.* Two-dimensional ferroelectric channel transistors integrating ultra-fast memory and neural computing. *Nature Communications* **12**, 53 (2021). https://doi.org/10.1038/s41467-020-20257-2

10 Cheng, J. *et al.* 2D reconfigurable memory for integrated optical sensing and multifunctional image processing. *ACS Applied Materials & Interfaces* **17**, 25467–25477 (2025). https://doi.org/10.1021/acsami.5c01496

11 Sui, F. *et al.* Sliding ferroelectricity in van der Waals layered γ-InSe semiconductor. *Nature Communications* **14**, 36 (2023). https://doi.org/10.1038/s41467-022-35490-0

12 Ding, W. *et al.* Prediction of intrinsic two-dimensional ferroelectrics in In2Se3 and other III2-VI3 van der Waals materials. *Nature Communications* **8**, 14956 (2017). https://doi.org/10.1038/ncomms14956

13 Tao, X. & Gu, Y. Crystalline–crystalline phase transformation in two-dimensional In2Se3 thin layers. *Nano Letters* **13**, 3501–3505 (2013). https://doi.org/10.1021/nl400888p

14 Han, W. *et al.* Phase-controllable large-area two-dimensional In2Se3 and ferroelectric heterophase junction. *Nature Nanotechnology* **18**, 55–63 (2023). https://doi.org/10.1038/s41565-022-01257-3

15 Song, S. *et al.* Wafer-scale growth of two-dimensional, phase-pure InSe. *Matter* **6**, 3483–3498 (2023). https://doi.org/10.1016/j.matt.2023.07.012

16 Lee, J. *et al.* Wafer-scale uniform non-ferroelectric κ-phase In2Se3 transistors. *Nature*

*Communications* **17**, 5901 (2026). https://doi.org/10.1038/s41467-026-72553-y
17 Zheng, C. *et al.* Room temperature in-plane ferroelectricity in van der Waals In2Se3. *Science Advances* **4**, eaar7720 (2018). https://doi.org/10.1126/sciadv.aar7720
18 Xu, C. *et al.* Two-dimensional antiferroelectricity in nanostripe-ordered In2Se3. *Physical Review Letters* **125**, 047601 (2020). https://doi.org/10.1103/PhysRevLett.125.047601
19 Lee, S. *et al.* Low-temperature processed beta-phase In2Se3 ferroelectric semiconductor thin film transistors. *2D Materials* **9**, 025023 (2022). https://doi.org/10.1088/2053-1583/ac5b17
20 Athena, F. F. *et al.* A van der Waals material exhibiting room temperature broken inversion symmetry with ferroelectricity. *Advanced Science* **13**, e18784 (2026). https://doi.org/10.1002/advs.202518784
21 Yang, M. *et al.* Structural and phase engineering in two-dimensional polymorphic In2Se3. *Chinese Physics Letters* **42**, 120705 (2025). https://doi.org/10.1088/0256-307X/42/12/120705
22 Kimoto, K. *et al.* Local crystal structure analysis with several picometer precision using scanning transmission electron microscopy. *Ultramicroscopy* **110**, 778–782 (2010). https://doi.org/10.1016/j.ultramic.2009.11.014
23 Yankovich, A. B. *et al.* Picometre-precision analysis of scanning transmission electron microscopy images of platinum nanocatalysts. *Nature Communications* **5**, 4155 (2014). https://doi.org/10.1038/ncomms5155
24 Gupta, P. *et al.* Perpendicular switching of polarization in layered ferroelectrics. *Nature* **656**, 67–72 (2026). https://doi.org/10.1038/s41586-026-10839-3
25 Wu, Y. *et al.* Stacking selected polarization switching and phase transition in vdW ferroelectric α-In2Se3 junction devices. *Nature Communications* **15**, 10481 (2024). https://doi.org/10.1038/s41467-024-54841-7
26 Buragohain, P. *et al.* Quantification of the electromechanical measurements by piezoresponse force microscopy. *Advanced Materials* **34**, 2206237 (2022). https://doi.org/10.1002/adma.202206237
27 Jesse, S., Baddorf, A. P. & Kalinin, S. V. Switching spectroscopy piezoresponse force microscopy of ferroelectric materials. *Applied Physics Letters* **88**, 062908 (2006). https://doi.org/10.1063/1.2172216
28 Gruverman, A. *et al.* Spatial inhomogeneity of imprint and switching behavior in ferroelectric capacitors. *Applied Physics Letters* **82**, 3071–3073 (2003). https://doi.org/10.1063/1.1570942

29 Cho, H. *et al.* Write cycling endurance exceeding $10^{10}$ in sub-50 nm ferroelectric AlScN. *Nature Communications* **17**, 1507 (2026). https://doi.org/10.1038/s41467-025-68221-2

30 Wang, D. *et al.* Sub-microsecond polarization switching in (Al,Sc)N ferroelectric capacitors grown on complementary metal–oxide–semiconductor-compatible aluminum electrodes. *physica status solidi (RRL) – Rapid Research Letters* **15**, 2000575 (2021). https://doi.org/10.1002/pssr.202000575

31 Robertson, J. & Wallace, R. M. High-K materials and metal gates for CMOS applications. *Materials Science and Engineering: R: Reports* **88**, 1–41 (2015). https://doi.org/10.1016/j.mser.2014.11.001

32 Wu, D. *et al.* Thickness-dependent dielectric constant of few-layer In2Se3 nanoflakes. *Nano Letters* **15**, 8136–8140 (2015). https://doi.org/10.1021/acs.nanolett.5b03575

33 Yasuda, K. *et al.* Ultrafast high-endurance memory based on sliding ferroelectrics. *Science* **385**, 53–56 (2024). https://doi.org/10.1126/science.adp3575

34 Fan, A. *et al.* Tailored sliding ferroelectricity for ultrahigh fatigue resistance in stacked trilayer MoS2 crystals. *Science Advances* **11**, eadx8192 (2025). https://doi.org/10.1126/sciadv.adx8192

35 Poh, S. M. *et al.* Molecular-beam epitaxy of two-dimensional In2Se3 and its giant electroresistance switching in ferroresistive memory junction. *Nano Letters* **18**, 6340–6346 (2018). https://doi.org/10.1021/acs.nanolett.8b02688

36 Wei, Y. *et al.* A rhombohedral ferroelectric phase in epitaxially strained Hf0.5Zr0.5O2 thin films. *Nature Materials* **17**, 1095–1100 (2018). https://doi.org/10.1038/s41563-018-0196-0

37 Kim, S. J. *et al.* Low-voltage operation and high endurance of 5-nm ferroelectric Hf0.5Zr0.5O2 capacitors. *Applied Physics Letters* **113**, 182903 (2018). https://doi.org/10.1063/1.5052012

38 Böscke, T. S., Müller, J., Braeuhaus, D., Schröder, U. & Böttger, U. Ferroelectricity in hafnium oxide thin films. *Applied Physics Letters* **99**, 102903 (2011). https://doi.org/10.1063/1.3634052

39 Chen, K.-T. *et al.* Improvement on ferroelectricity and endurance of ultra-thin HfZrO2 capacitor with molybdenum capping electrode. *2019 Electron Devices*

*Technology and Manufacturing Conference (EDTM)*, 62–64 (2019). https://doi.org/10.1109/EDTM.2019.8731272

40 Cheema, S. S. *et al.* Enhanced ferroelectricity in ultrathin films grown directly on silicon. *Nature* **580**, 478–482 (2020). https://doi.org/10.1038/s41586-020-2208-x

41 Li, Y.-C. *et al.* Domain switching characteristics in Ga-doped HfO2 ferroelectric thin films with low coercive field. *Nano Letters* **24**, 6585–6591 (2024). https://doi.org/10.1021/acs.nanolett.4c00263

42 Yun, Y. *et al.* Intrinsic ferroelectricity in Y-doped HfO2 thin films. *Nature Materials* **21**, 903–909 (2022). https://doi.org/10.1038/s41563-022-01282-6

43 Jiang, T. *et al.* Ferroelectricity in six-ångström-thick two-dimensional gallium oxide. *Nature Electronics* **9**, 1007–1017 (2026). https://doi.org/10.1038/s41928-026-01694-1

44 Wang, Y. *et al.* A stable rhombohedral phase in ferroelectric Hf(Zr)1+xO2 capacitor with ultralow coercive field. *Science* **381**, 558–563 (2023). https://doi.org/10.1126/science.adf6137

45 Jo, J., Kim, Y., Noh, T., Yoon, J.-G. & Song, T. Coercive fields in ultrathin BaTiO3 capacitors. *Applied Physics Letters* **89**, 232909 (2006). https://doi.org/10.1063/1.2402238

46 Feng, G. *et al.* A ferroelectric fin diode for robust non-volatile memory. *Nature Communications* **15**, 513 (2024). https://doi.org/10.1038/s41467-024-44759-5

47 Rodríguez Contreras, J., Kohlstedt, H., Poppe, U., Waser, R. & Buchal, C. Surface treatment effects on the thickness dependence of the remanent polarization of PbZr0.52Ti0.48O3 capacitors. *Applied Physics Letters* **83**, 126–128 (2003). https://doi.org/10.1063/1.1590431

48 Kim, K.-H. *et al.* Scalable CMOS back-end-of-line-compatible AlScN/two-dimensional channel ferroelectric field-effect transistors. *Nature Nanotechnology* **18**, 1044–1050 (2023). https://doi.org/10.1038/s41565-023-01399-y

49 He, Y. *et al.* Metal-ferroelectric AlScN-semiconductor memory devices on SiC wafers. *Applied Physics Letters* **123**, 122901 (2023). https://doi.org/10.1063/5.0161746

50 Hu, Z. *et al.* Compositional gradient engineering for enhanced ferroelectricity in ultrathin AlScN. Preprint at https://doi.org/10.48550/arXiv.2606.12568 (2026).

51 Wang, D. *et al.* Thickness scaling down to 5 nm of ferroelectric ScAlN on CMOS compatible molybdenum grown by molecular beam epitaxy. *Applied Physics Letters*

**122**, 052101 (2023). https://doi.org/10.1063/5.0136265

52 Wan, S. *et al.* Nonvolatile ferroelectric memory effect in ultrathin α-In2Se3. *Advanced Functional Materials* **29**, 1808606 (2019). https://doi.org/10.1002/adfm.201808606

53 Wu, J. *et al.* High tunnelling electroresistance in a ferroelectric van der Waals heterojunction via giant barrier height modulation. *Nature Electronics* **3**, 466–472 (2020). https://doi.org/10.1038/s41928-020-0441-9

54 de la Barrera, S. C. *et al.* Direct measurement of ferroelectric polarization in a tunable semimetal. *Nature Communications* **12**, 5298 (2021). https://doi.org/10.1038/s41467-021-25587-3

55 Long, Y. *et al.* Competing ferroelectric polarization and defect migration induced resistive switching in β′-In2Se3. *Nano Letters* **25**, 2922–2930 (2025). https://doi.org/10.1021/acs.nanolett.4c06314

56 Aruchamy, N. *et al.* Influence of tensile vs. compressive stress on fatigue of lead zirconate titanate thin films. *Journal of the European Ceramic Society* **41**, 6991–6999 (2021). https://doi.org/10.1016/j.jeurceramsoc.2021.07.010

57 Cheng, Y. *et al.* Reversible transition between the polar and antipolar phases and its implications for wake-up and fatigue in HfO2-based ferroelectric thin film. *Nature Communications* **13**, 645 (2022). https://doi.org/10.1038/s41467-022-28236-5

58 Tan, W. L., Faber, K. T. & Kochmann, D. M. In-situ observation of evolving microstructural damage and associated effective electro-mechanical properties of PZT during bipolar electrical fatigue. *Acta Materialia* **164**, 704–713 (2019). https://doi.org/10.1016/j.actamat.2018.10.065

59 Wang, X. *et al.* Van der Waals engineering of ferroelectric heterostructures for long-retention memory. *Nature Communications* **12**, 1109 (2021). https://doi.org/10.1038/s41467-021-21320-2

60 Jiang, Y. *et al.* Enabling ultra-low-voltage switching in BaTiO3. *Nature Materials* **21**, 779–785 (2022). https://doi.org/10.1038/s41563-022-01266-6

61 Wang, W. *et al.* A ferroelectric capacitor with an asymmetric double-layered ferroelectric structure comprising a liquid-delivery MOCVD Pb(Zr, Ti)O3 layer and a sputter-deposited La-doped Pb(Zr, Ti)O3 for highly reliable FeRAM. *Applied Physics Letters* **125**, 192901 (2024). https://doi.org/10.1063/5.0230646

62 Yang, T. H. *et al.* Ferroelectric transistors based on shear-transformation-mediated rhombohedral-stacked molybdenum disulfide. *Nature Electronics* **7**, 29–38 (2024). https://doi.org/10.1038/s41928-023-01073-0

63 Baek, S. *et al.* Ferroelectric field-effect-transistor integrated with ferroelectrics heterostructure. *Advanced Science* **9**, 2200566 (2022). https://doi.org/10.1002/advs.202200566

64 Huang, W. *et al.* Gate-coupling-enabled robust hysteresis for nonvolatile memory and programmable rectifier in van der Waals ferroelectric heterojunctions. *Advanced Materials* **32**, 1908040 (2020). https://doi.org/10.1002/adma.201908040

65 Li, W. *et al.* Emergence of ferroelectricity in a nonferroelectric monolayer. *Nature Communications* **14**, 2757 (2023). https://doi.org/10.1038/s41467-023-38445-1

66 Shenai, D. V., Timmons, M. L., DiCarlo, R. L., Lemnah, G. K. & Stennick, R. S. Correlation of vapor pressure equation and film properties with trimethylindium purity for the MOVPE grown III–V compounds. *Journal of Crystal Growth* **248**, 91–98 (2003). https://doi.org/10.1016/S0022-0248(02)01854-7

67 Karlson, U., Frankenberger Jr, W. & Spencer, W. Physicochemical properties of dimethyl selenide and dimethyl diselenide. *Journal of Chemical and Engineering Data* **39**, 608–610 (1994). https://doi.org/10.1021/je00015a049

68 Jones, L. *et al.* Smart Align—a new tool for robust non-rigid registration of scanning microscope data. *Advanced Structural and Chemical Imaging* **1**, 8 (2015). https://doi.org/10.1186/s40679-015-0008-4

69 Sang, X. & LeBeau, J. M. Revolving scanning transmission electron microscopy: Correcting sample drift distortion without prior knowledge. *Ultramicroscopy* **138**, 28–35 (2014). https://doi.org/10.1016/j.ultramic.2013.12.004

70 Ophus, C., Ciston, J. & Nelson, C. T. Correcting nonlinear drift distortion of scanning probe and scanning transmission electron microscopies from image pairs with orthogonal scan directions. *Ultramicroscopy* **162**, 1–9 (2016). https://doi.org/10.1016/j.ultramic.2015.12.002

71 Koch, C. T. *Determination of Core Structure Periodicity and Point Defect Density along Dislocations*. PhD thesis, Arizona State University (2002).

72 Kim, Y.-M. *et al.* Probing oxygen vacancy concentration and homogeneity in solid-oxide fuel-cell cathode materials on the subunit-cell level. *Nature Materials* **11**, 888–894 (2012). https://doi.org/10.1038/nmat3393

73 Kim, Y.-M. *et al.* Direct observation of ferroelectric field effect and vacancy-controlled screening at the BiFeO3/LaxSr1−xMnO3 interface. *Nature Materials* **13**, 1019–1025 (2014). https://doi.org/10.1038/nmat4058

74 de la Peña, F. *et al.* Hyperspy/hyperspy: V2.4.0. *Zenodo* (2026). https://doi.org/10.5281/zenodo.18379337

75 Nord, M., Vullum, P. E., MacLaren, I., Tybell, T. & Holmestad, R. Atomap: a new software tool for the automated analysis of atomic resolution images using two-dimensional Gaussian fitting. *Advanced Structural and Chemical Imaging* **3**, 9 (2017). https://doi.org/10.1186/s40679-017-0042-5

76 Kim, Y.-M. *et al.* Interplay of octahedral tilts and polar order in BiFeO3 films. *Advanced Materials* **25**, 2497–2504 (2013). https://doi.org/10.1002/adma.201204584

77 Min, T. *et al.* Cooperative evolution of polar distortion and nonpolar rotation of oxygen octahedra in oxide heterostructures. *Science Advances* **7**, eabe9053 (2021). https://doi.org/10.1126/sciadv.abe9053

78 Lee, H. *et al.* Hybrid ferroelectric tunnel junctions with intrinsic nonlinearity and self-rectification via interfacial reconstruction. *Advanced Functional Materials*, e78527 (2026). https://doi.org/10.1002/adfm.78527

79 Kresse, G. & Furthmüller, J. Efficient iterative schemes for ab initio total-energy calculations using a plane-wave basis set. *Physical Review B* **54**, 11169 (1996). https://doi.org/10.1103/PhysRevB.54.11169

80 Blöchl, P. E. Projector augmented-wave method. *Physical Review B* **50**, 17953 (1994). https://doi.org/10.1103/PhysRevB.50.17953

81 Kresse, G. & Joubert, D. From ultrasoft pseudopotentials to the projector augmented-wave method. *Physical Review B* **59**, 1758 (1999). https://doi.org/10.1103/PhysRevB.59.1758

82 Heyd, J., Scuseria, G. E. & Ernzerhof, M. Hybrid functionals based on a screened Coulomb potential. *The Journal of Chemical Physics* **118**, 8207–8215 (2003). https://doi.org/10.1063/1.1564060

83 Krukau, A. V., Vydrov, O. A., Izmaylov, A. F. & Scuseria, G. E. Influence of the exchange screening parameter on the performance of screened hybrid functionals. *The Journal of Chemical Physics* **125**, 224106 (2006). https://doi.org/10.1063/1.2404663

84 Caldeweyher, E. *et al.* A generally applicable atomic-charge dependent London dispersion correction. *The Journal of Chemical Physics* **150**, 154122 (2019). https://doi.org/10.1063/1.5090222

85 Caldeweyher, E., Mewes, J.-M., Ehlert, S. & Grimme, S. Extension and evaluation of the D4 London-dispersion model for periodic systems. *Physical Chemistry Chemical Physics* **22**, 8499–8512 (2020). https://doi.org/10.1039/D0CP00502A

86 King-Smith, R. & Vanderbilt, D. Theory of polarization of crystalline solids. *Physical Review B* **47**, 1651 (1993). https://doi.org/10.1103/PhysRevB.47.1651

87 Resta, R. Macroscopic polarization in crystalline dielectrics: the geometric phase approach. *Reviews of Modern Physics* **66**, 899 (1994). https://doi.org/10.1103/RevModPhys.66.899

88 Hinuma, Y., Pizzi, G., Kumagai, Y., Oba, F. & Tanaka, I. Band structure diagram paths based on crystallography. *Computational Materials Science* **128**, 140–184 (2017). https://doi.org/10.1016/j.commatsci.2016.10.015

# Unveiling ferroelectricity in wafer-scale beta-phase $In_2Se_3$

*Kyungmin Yang[a,+], Hyunmin Cho[a,+], Jihwan Jeon[b,c,+], Tae Hyeon Park[b,c], Seojeong Ro[c], Hye Jung Kim[b], Jung-Hoon Lee[d,e], Jaewon Wang[a], Hyeon Suk Shin[b,c,f], Young-Min Kim[b,c], Chang Woo Myung[b,c], Roy H. Olsson III[a], Seunguk Song[b,c,*], Deep Jariwala[a,*]*

[a] Department of Electrical and Systems Engineering, University of Pennsylvania, Philadelphia, PA 19104, USA

[b] Center for 2D Quantum Heterostructures (2DQH), Institute for Basic Science (IBS), Suwon, 16419, Republic of Korea

[c] Department of Energy Science, Sungkyunkwan University (SKKU), Suwon, 16419, Republic of Korea

[d] Computational Science Research Center, Korea Institute of Science and Technology (KIST), Seoul, 02792, Republic of Korea

[e] KU-KIST Graduate School of Converging Science and Technology, Korea University, Seoul, 02841, Republic of Korea

[f] Department of Chemistry, Sungkyunkwan University (SKKU), Suwon, 16419, Republic of Korea

[+] These authors contributed equally

* E-mail: seunguk@skku.edu, dmj@engineering.upenn.edu

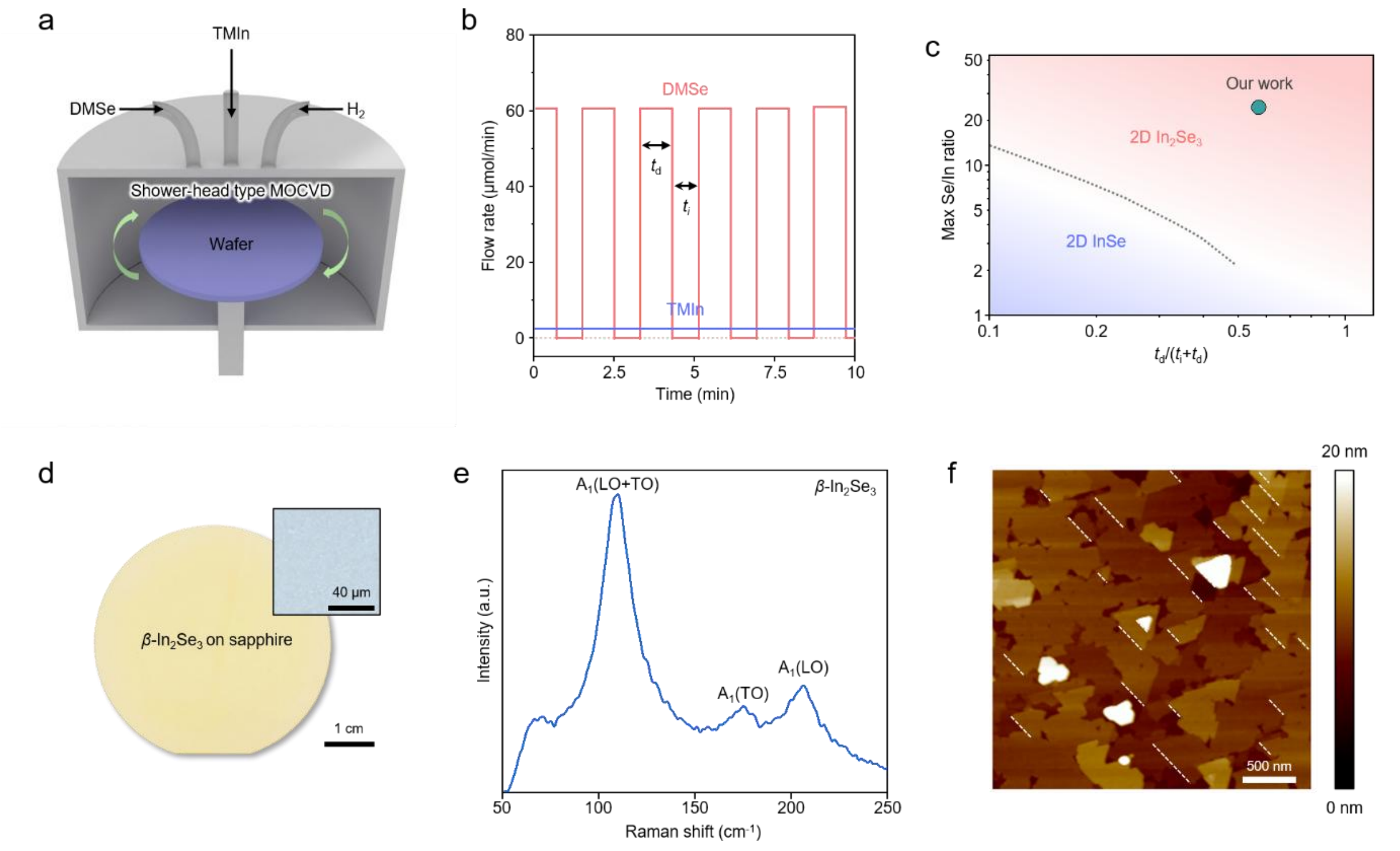

**Supplementary Figure 1.** (a) Schematic illustration of the showerhead-type MOCVD system used for the growth of $In_2Se_3$. (b) Source deposition time and flow rate of DMSe and TMIn during the MOCVD growth process. (c) Se/In ratio as a function of $t_d/(t_i + t_d)$, where $t_i$ and $t_d$ are the TMIn and DMSe delivery durations within one precursor cycle, indicating the growth regime used in this work for forming 2D $In_2Se_3$. The films in this work were grown with $t_i$ = 45 s and $t_d$ = 60 s ($t_d/(t_i + t_d) = 0.57$) at a maximum Se/In ratio of ~24 (Methods). (d) Photograph of 2-inch β-$In_2Se_3$ grown by MOCVD on sapphire wafer with its optical microscopy image (inset). (e) Raman spectra of MOCVD-grown β-$In_2Se_3$ film. (f) AFM topography image of the β-$In_2Se_3$ film on sapphire. The dashed lines (white) represent the edges of the $In_2Se_3$ domains, revealing crystal orientation.

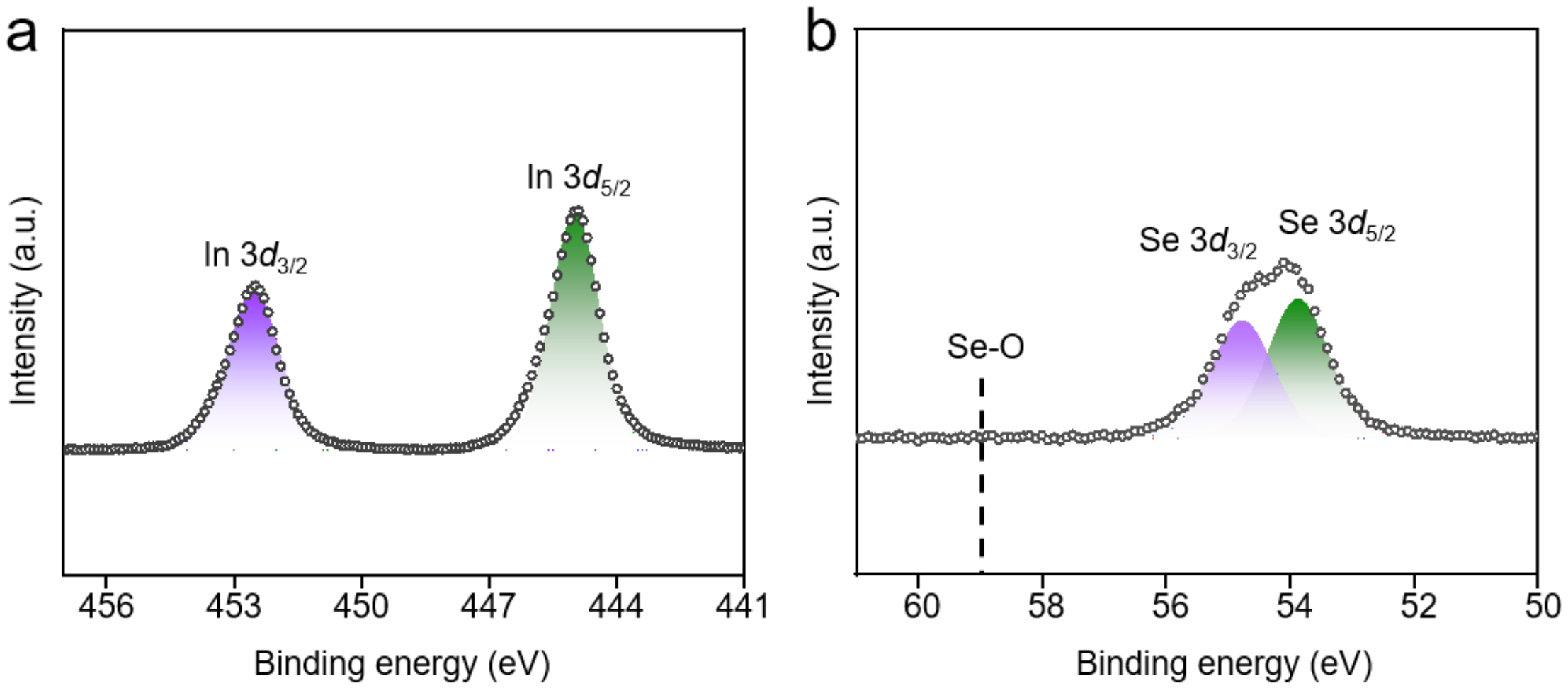

**Supplementary Figure 2.** XPS spectra showing (a) In 3d and (b) Se 3d core levels, with no detectable oxide-related components. The Se/In ratio obtained from the integrated core-level intensities is ≈1.47; because the accuracy of XPS quantification is about ±2%, the ratio is used only as a consistency check for the phase.

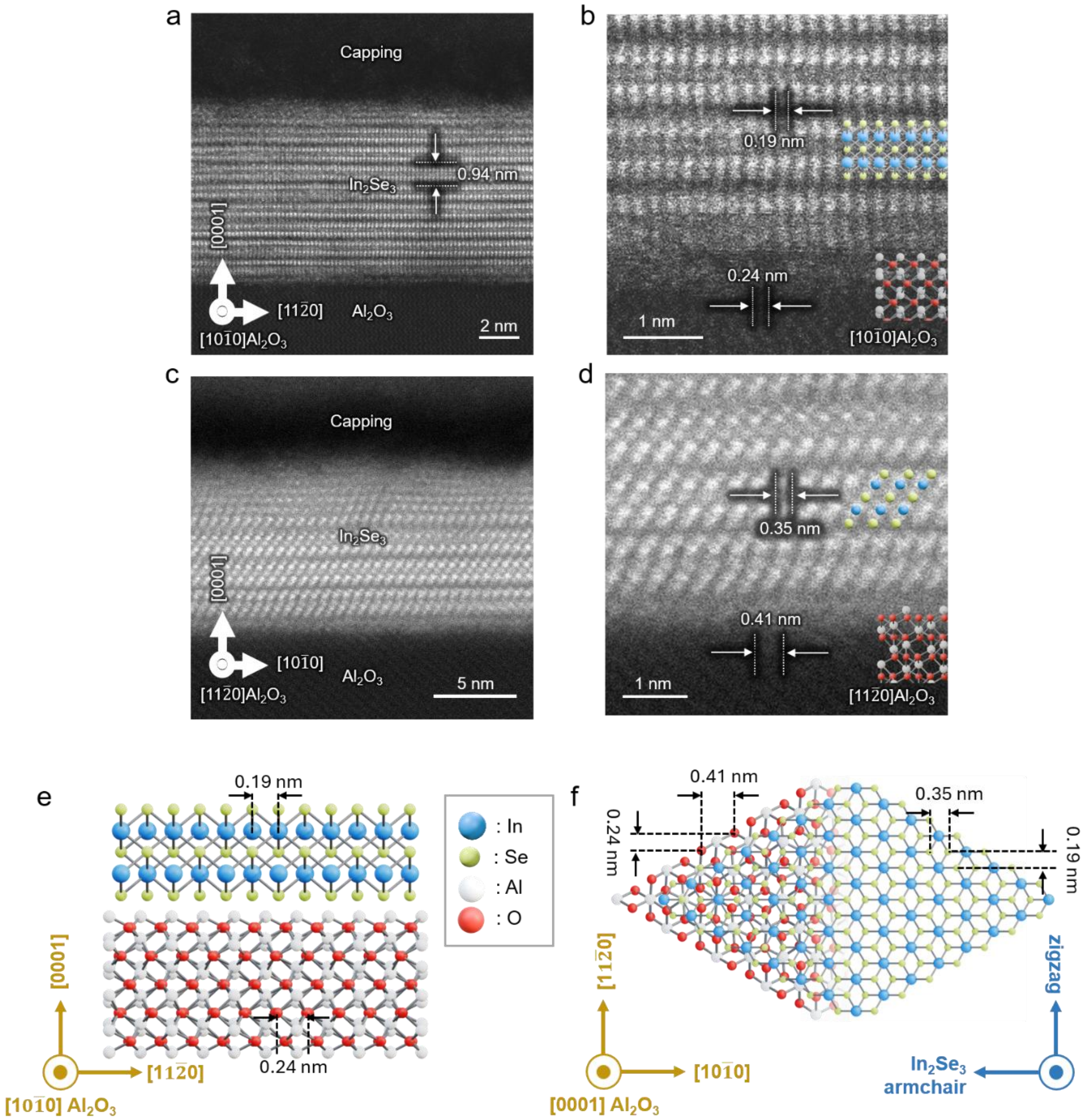

**Supplementary Figure 3.** (a, c) Cross-sectional scanning transmission electron microscopy (STEM) images of the few-layer $In_2Se_3$ film along the (a) $[10\bar{1}0]$ and (c) $[11\bar{2}0]$ zone axes of $Al_2O_3$. (b, d) Atomic-resolution STEM images of the $In_2Se_3$/$Al_2O_3$ interface along the (b) $[10\bar{1}0]$ and (d) $[11\bar{2}0]$ zone axes of $Al_2O_3$. Schematics of the crystal structure for $In_2Se_3$ and $Al_2O_3$ along the appropriate orientations are included as insets in (b) and (d). (e, f) Cross-section (e) and plan-view (f) schematics illustrating the epitaxial relationship between the $In_2Se_3$ film and the sapphire (0001) surface.

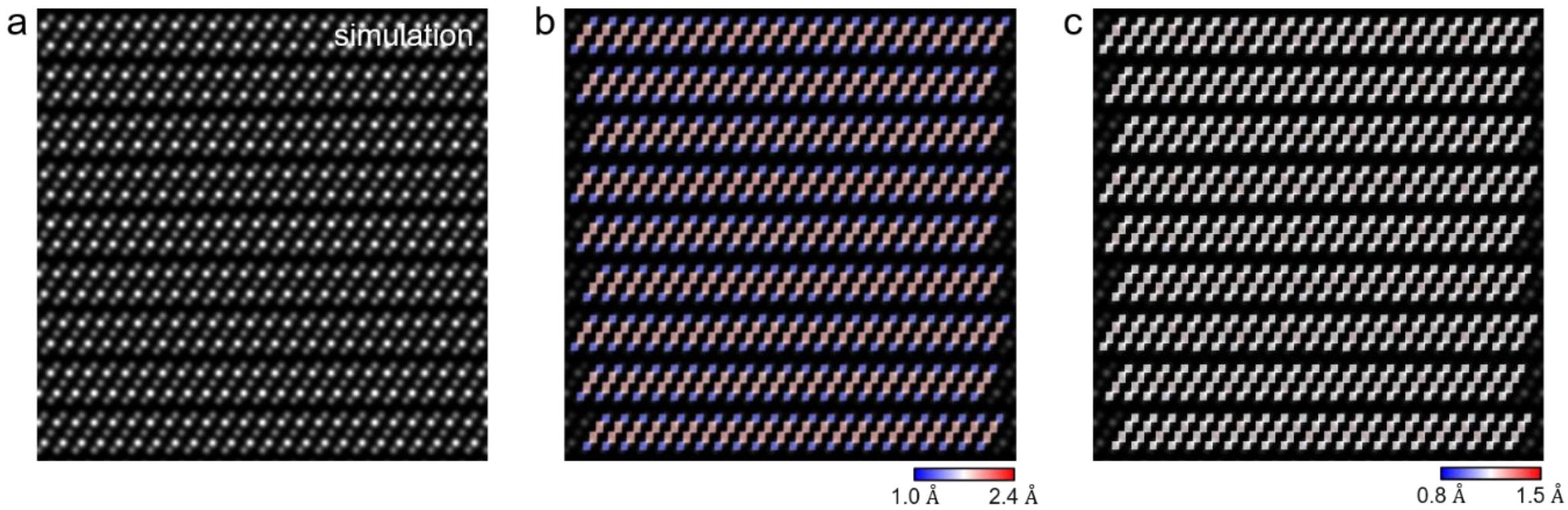

**Supplementary Figure 4.** Simulated HAADF-STEM image of β-$In_2Se_3$ and the corresponding projected atomic distance (PAD) maps. (a) HAADF-STEM image of the β-$In_2Se_3$ quintuple layers (QLs) simulated with qSTEM for the $[11\bar{2}0]$ projection using the DFT-relaxed centrosymmetric structure as the input model. The simulation conditions are described in Methods. (b) Out-of-plane PAD map obtained from (a). Within each Se1–In1–$Se_C$–In2–Se2 sequence constituting a QL, the out-of-plane component of the distance between adjacent atomic columns was measured and assigned, as a false color, to the rectangle whose opposite corners are the centers of the two columns. Se1 and Se2 denote the bottom and top Se, In1 and In2 the lower and upper In, and $Se_C$ the central Se. The color scale (1.0–2.4 Å) is the same as in Supplementary Figures 5 and 6. (c) In-plane PAD map obtained from (a), measured and displayed as in (b). The color scale (0.8–1.5 Å) is the same as in Supplementary Figures 7 and 8. In (b) and (c), every sequence shows the same color pattern, indicating that the analysis procedure itself introduces no appreciable spread in the PADs. The simulated image therefore serves as a reference against which the width of the distance distributions in the experimental images (Supplementary Figures 5–8) can be attributed to scan distortion, noise, and genuine atomic displacements.

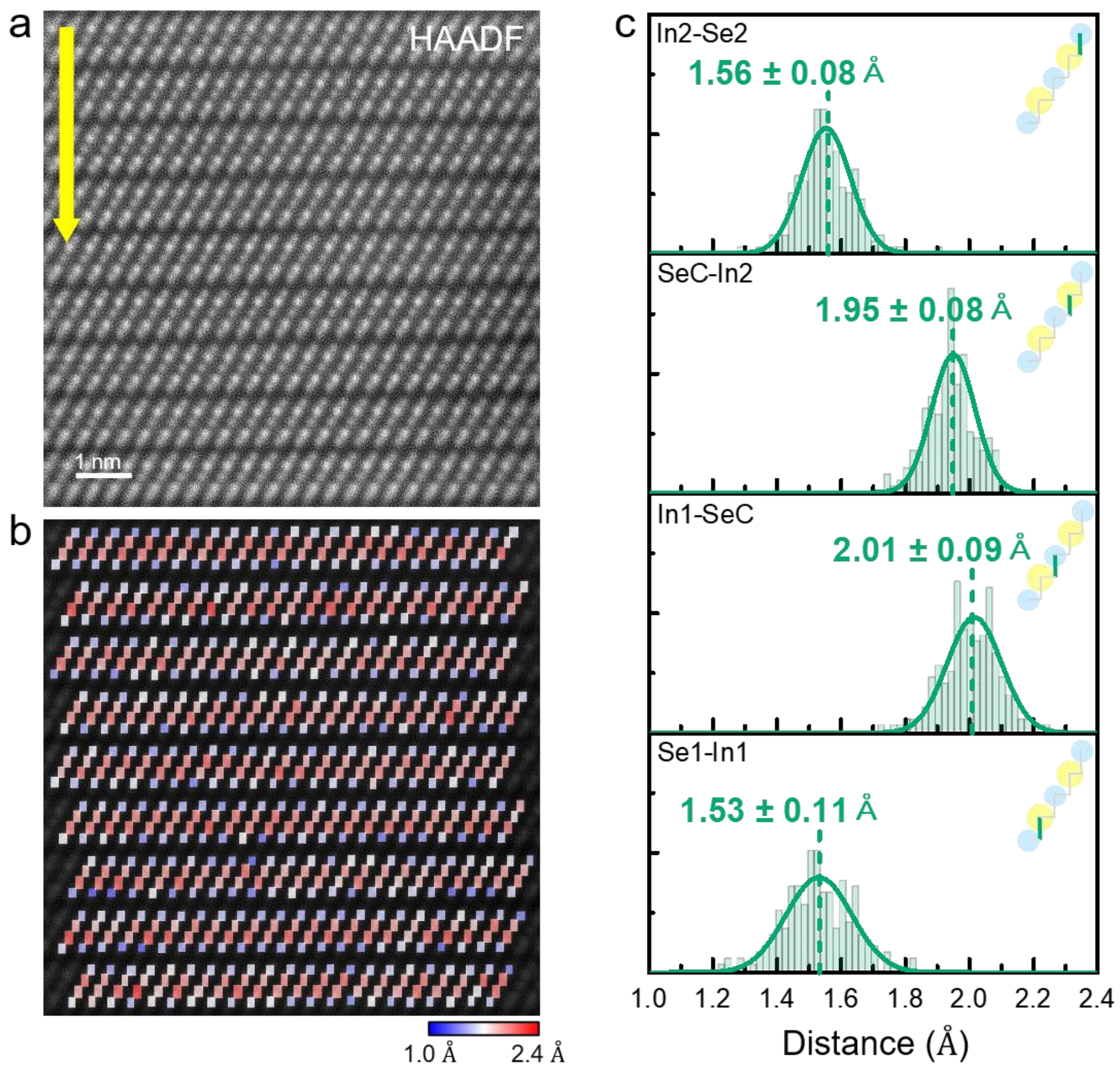

**Supplementary Figure 5.** Quantitative analysis of the out-of-plane atomic distances in β-$In_2Se_3$ in the Program state. (a) Cross-sectional HAADF-STEM image of a device switched to the Program state, viewed along the $[11\bar{2}0]$ direction. The yellow arrow indicates the fast-scan direction. (b) Out-of-plane projected atomic distance (PAD) map obtained from (a). Within each Se1–In1–$Se_C$–In2–Se2 sequence constituting a QL (atom labels as defined in Supplementary Figure 4b), the out-of-plane component of the distance between adjacent atomic columns was measured and assigned, as a false color, to the rectangle whose opposite corners are the centers of the two columns. The color scale spans 1.0–2.4 Å. (c) Histograms of the out-of-plane distances of the four consecutive bonds within the QL (Se1–In1, In1–$Se_C$, $Se_C$–In2, and In2–Se2) measured from (a) plotted as counts versus distance ($N$ = 206 sequences). Solid lines are Gaussian fits, dashed lines denote the mean values, and the values given in each panel are the mean ± standard deviation. The schematic at the upper right of each panel marks the atom pair of the measured distance with a green bar (light blue, Se; yellow, In). The difference between the In1–$Se_C$ and $Se_C$–In2 distances reflects the direction and magnitude of the displacement of the central Se from the midpoint between In1 and In2.

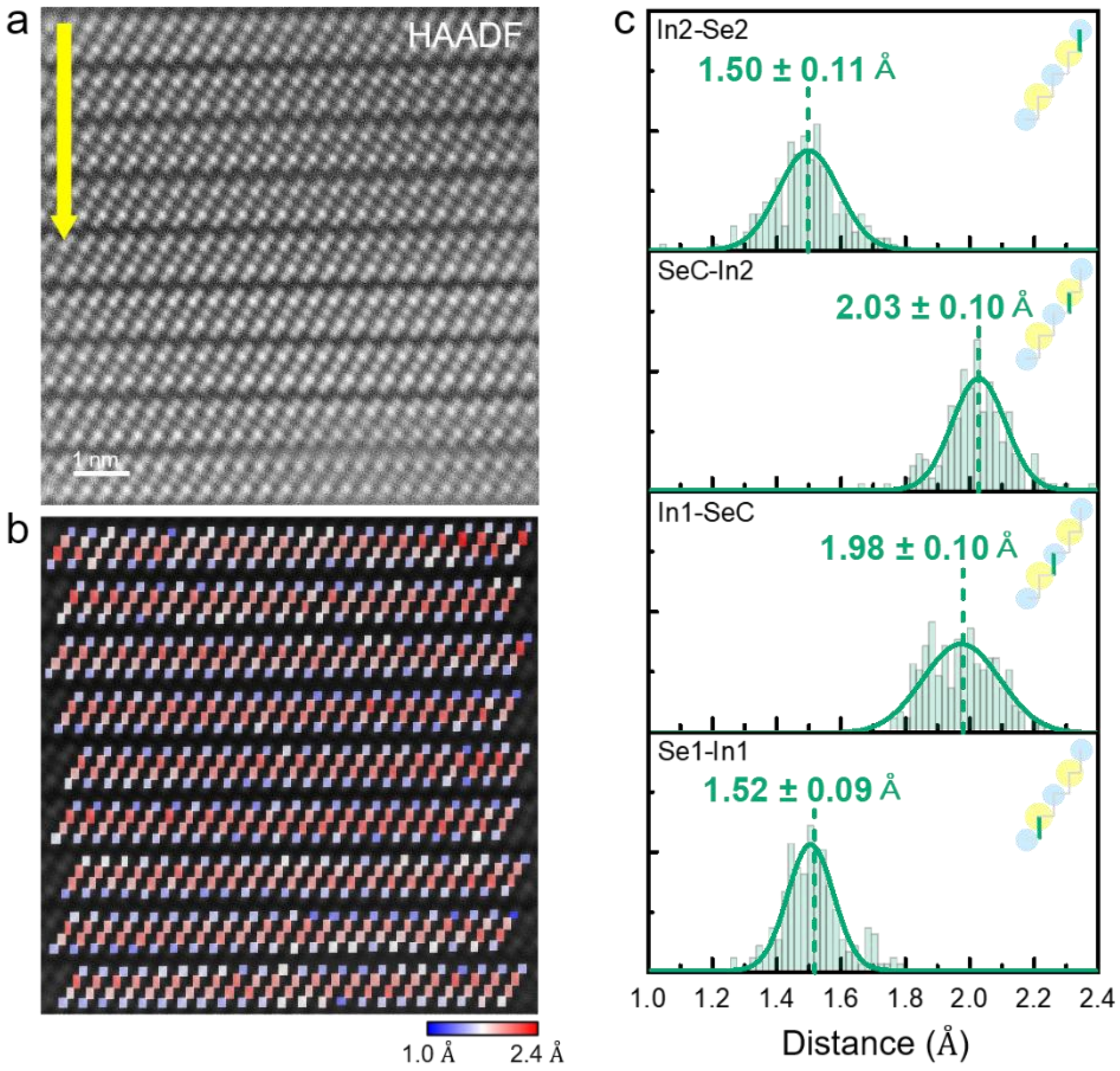

**Supplementary Figure 6.** Quantitative analysis of the out-of-plane atomic distances in β-$In_2Se_3$ in the Erase state. (a) Cross-sectional HAADF-STEM image of a device switched to the Erase state, viewed along the $[11\bar{2}0]$ direction. The yellow arrow indicates the fast-scan direction. (b) Out-of-plane PAD map obtained from (a), measured and displayed as in Supplementary Figure 5b (atom labels as defined in Supplementary Figure 4b). The color scale spans 1.0–2.4 Å. (c) Histograms of the out-of-plane distances of the four consecutive bonds measured from (a), plotted as counts versus distance ($N$ = 209 sequences). Solid lines are Gaussian fits, dashed lines denote the mean values, and the values given in each panel are the mean ± standard deviation. The schematics are as in Supplementary Figure 5c. Compared with the Program state (Supplementary Figure 5c), the order of the In1–$Se_C$ and $Se_C$–In2 distances is reversed, indicating that the central Se is displaced in the opposite direction.

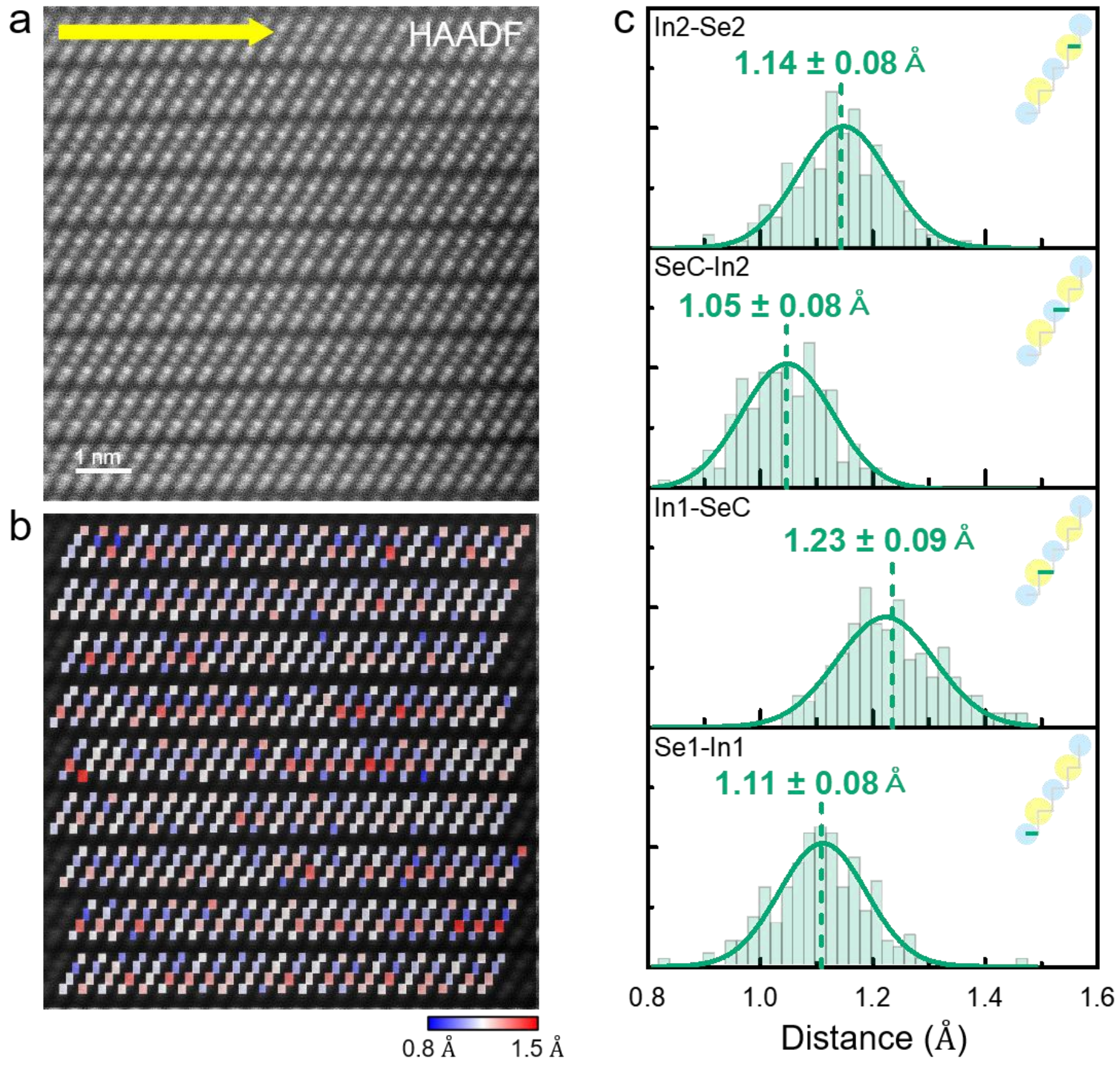

**Supplementary Figure 7.** Quantitative analysis of the in-plane atomic distances in β-$In_2Se_3$ in the Program state. (a) Cross-sectional HAADF-STEM image of a device switched to the Program state, viewed along the $[11\bar{2}0]$ direction. The yellow arrow indicates the fast-scan direction. (b) In-plane PAD map obtained from (a). Within each Se1–In1–$Se_C$–In2–Se2 sequence constituting a QL (atom labels as defined in Supplementary Figure 4b), the in-plane component of the distance between adjacent atomic columns was measured and assigned, as a false color, to the rectangle whose opposite corners are the centers of the two columns. The color scale spans 0.8–1.5 Å. (c) Histograms of the in-plane distances of the four consecutive bonds measured from (a), plotted as counts versus distance (N = 204 sequences). Solid lines are Gaussian fits, dashed lines denote the mean values, and the values given in each panel are the mean ± standard deviation. The schematics are as in Supplementary Figure 5c. The in-plane difference between the In1–$Se_C$ and $Se_C$–In2 distances (≈18 pm, which corresponds to a $Se_C$ offset of ≈9 pm) has the same sign in every QL, with no periodic sign alternation along the layers such as would be expected for the β′ nanostripe superstructure.

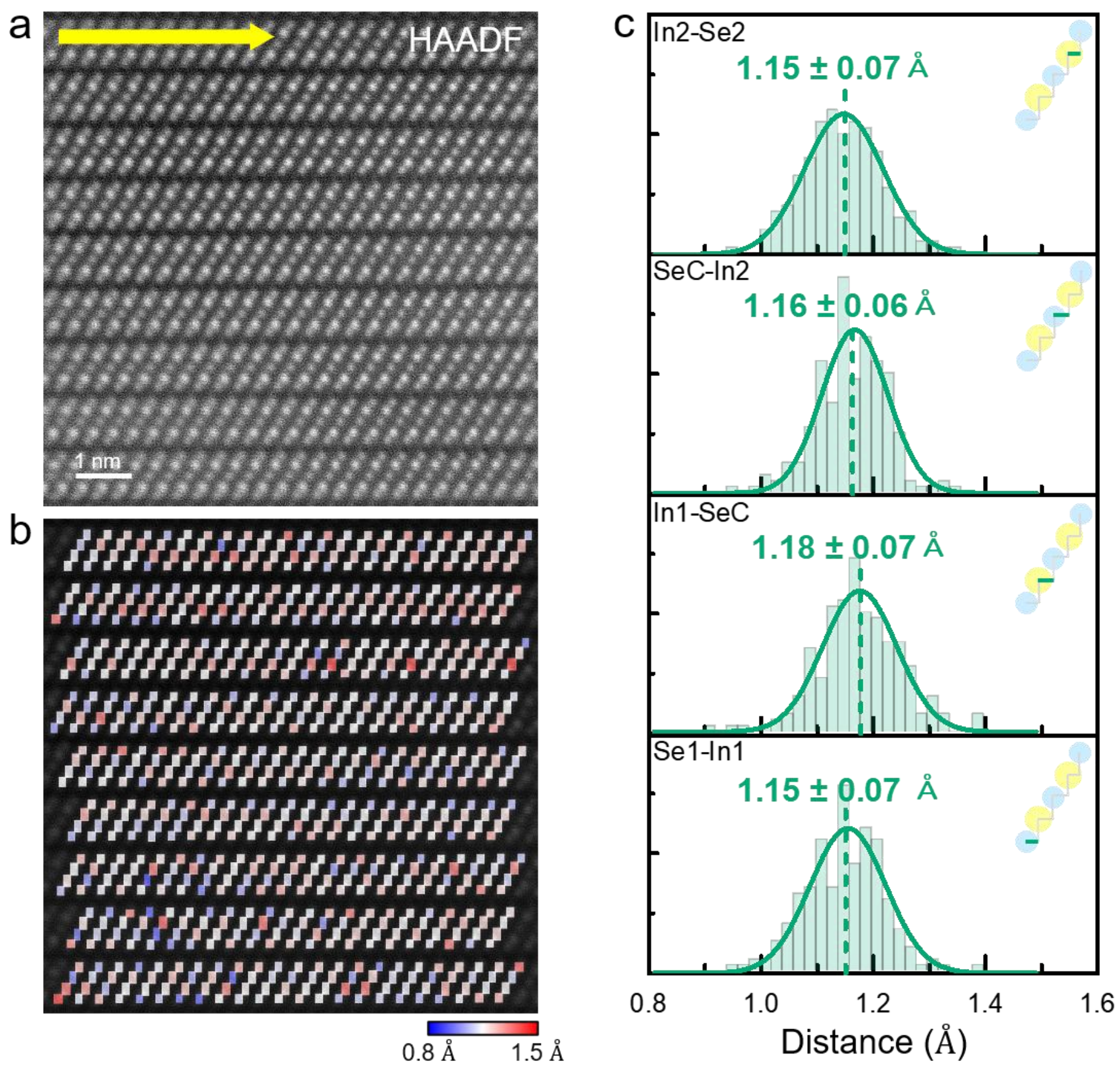

**Supplementary Figure 8.** Quantitative analysis of the in-plane atomic distances in β-$In_2Se_3$ in the Erase state. (a) Cross-sectional HAADF-STEM image of a device switched to the Erase state, viewed along the $[11\bar{2}0]$ direction. The yellow arrow indicates the fast-scan direction. (b) In-plane PAD map obtained from (a), measured and displayed as in Supplementary Figure 7b. The color scale spans 0.8–1.5 Å. (c) Histograms of the in-plane distances of the four consecutive bonds measured from (a), plotted as counts versus distance (N = 205 sequences). Solid lines are Gaussian fits, dashed lines denote the mean values, and the values given in each panel are the mean ± standard deviation. The schematics are as in Supplementary Figure 5c. In this lamella the in-plane In1–$Se_C$ and $Se_C$–In2 distances are equal within ≈2 pm. The in-plane component resolved in a cross-section depends on which in-plane variant of the polar structure the lamella samples, so that only the out-of-plane component is compared between the Program and Erase lamellae (main text).

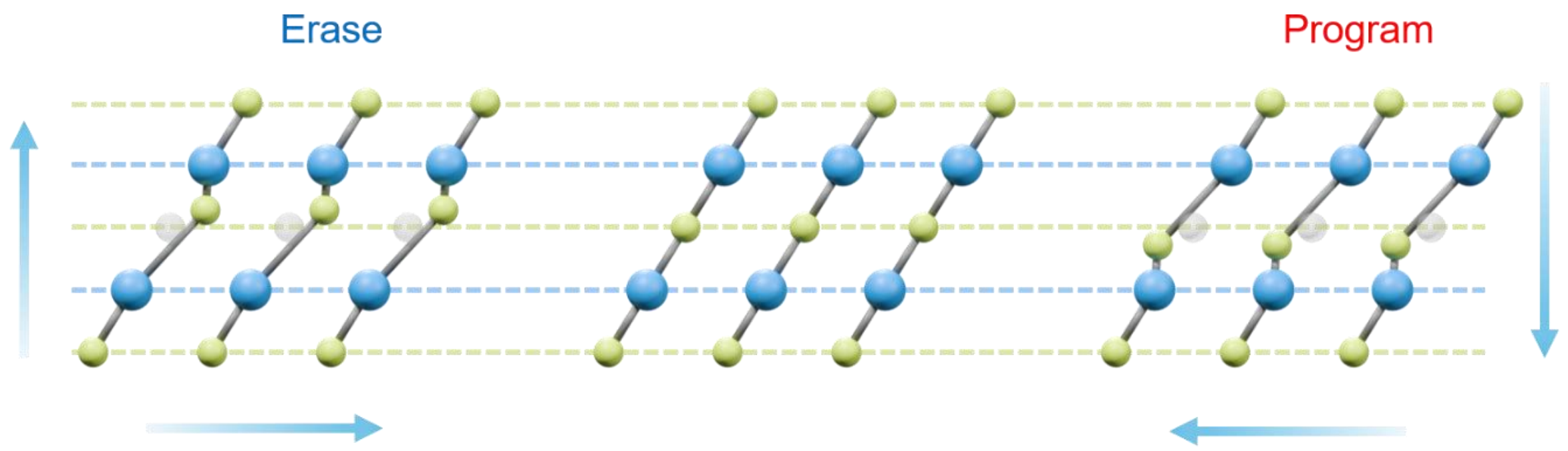

**Supplementary Figure 9.** Polar equilibrium structure of β-$In_2Se_3$ obtained from first-principles calculations (HSE06+D4). Erase and Program denote the two oppositely polarized states, which are related by inversion. Semi-transparent spheres mark the atomic positions of the centrosymmetric reference structure, and arrows indicate the out-of-plane and in-plane polarization components.

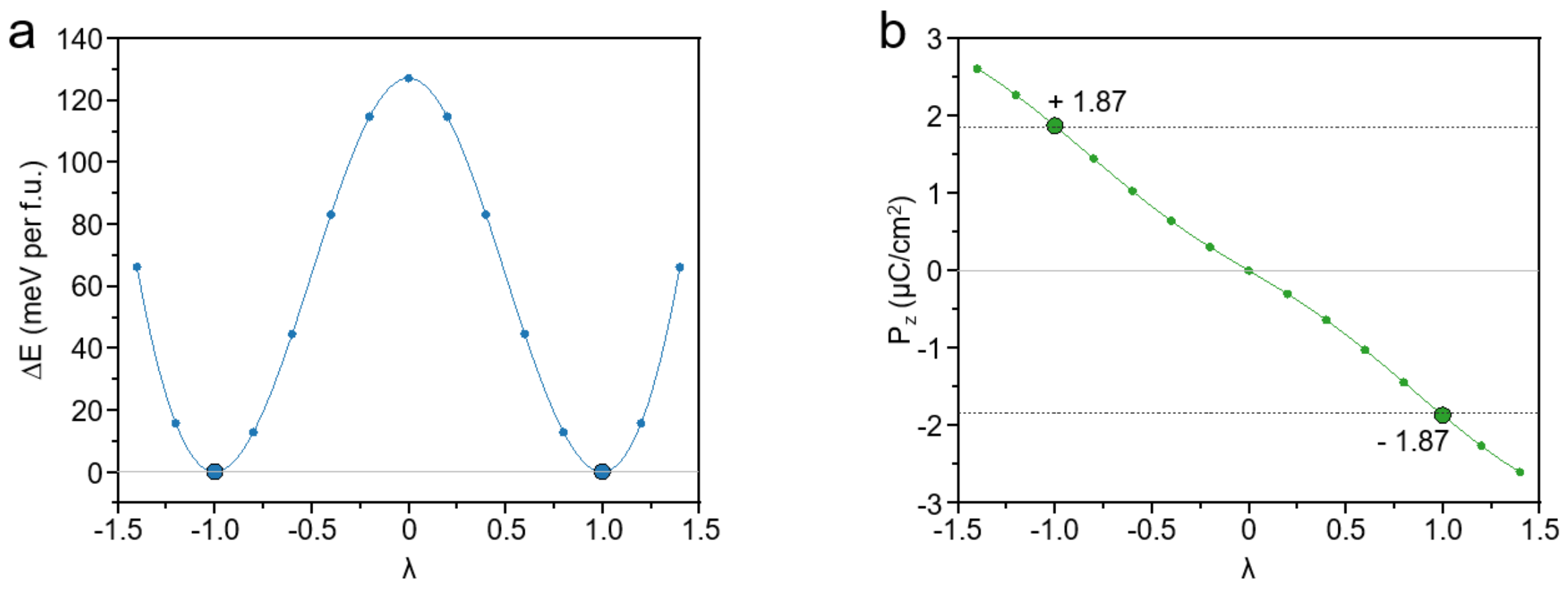

**Supplementary Figure 10.** Energy and polarization along the switching path connecting the two polar states of β-$In_2Se_3$. λ = ±1 correspond to the two polar states relaxed with HSE06+D4 (filled markers), λ = 0 to the centrosymmetric reference structure, and intermediate structures are obtained by linear interpolation of atomic coordinates. (a) Relative energy. The switching barrier is 127 meV per formula unit. (b) Out-of-plane polarization. The spontaneous polarization at λ = ±1 is ∓1.87 μC/$cm^2$. The in-plane component of the same polar mode is 21.1 μC/$cm^2$ at λ = ±1 and reverses along the same path.

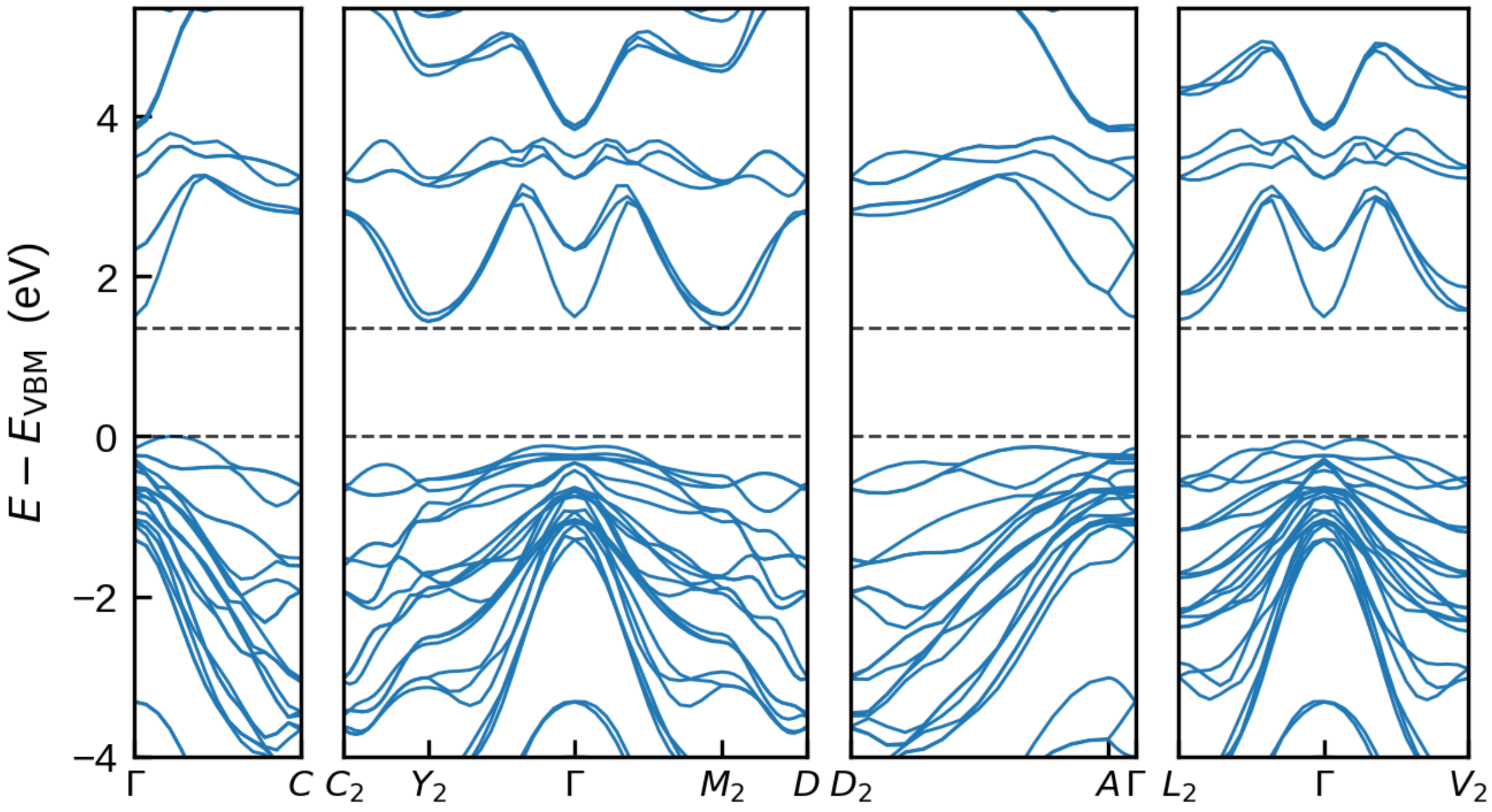

**Supplementary Figure 11.** Electronic band structure of the polar equilibrium structure of β-$In_2Se_3$ calculated with HSE06+D4. Energies are referenced to the valence band maximum, and the dashed lines mark the valence band maximum and the conduction band minimum. The band gap is 1.35 eV.

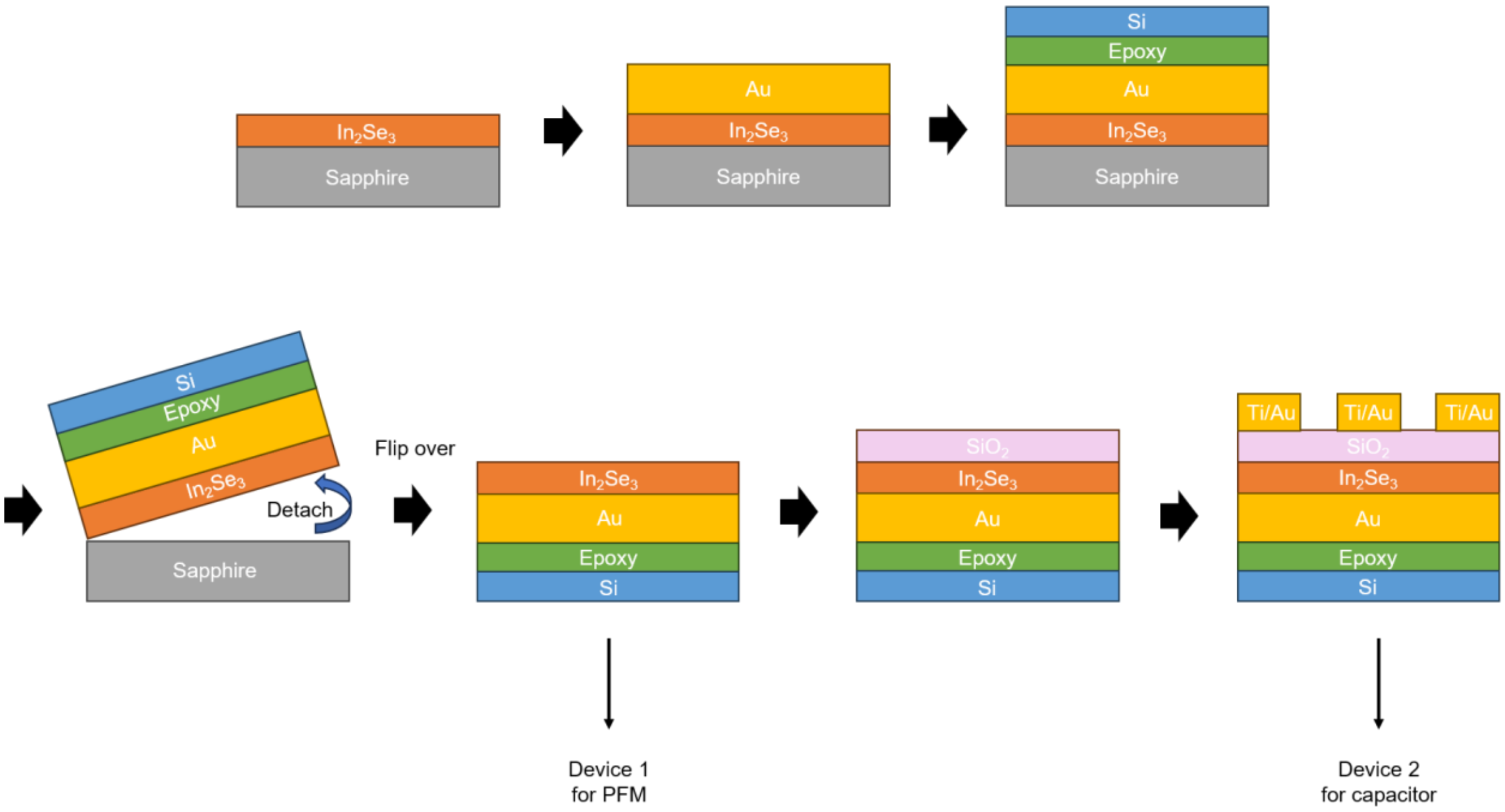

**Supplementary Figure 12.** Preparation of the Au-supported β-$In_2Se_3$ films used for PFM (Device 1) and for vertical capacitors (Device 2): Au deposition on the as-grown film, epoxy bonding to a Si carrier, mechanical detachment from the sapphire growth substrate, flipping, $SiO_2$ deposition by electron-beam evaporation and Ti/Au top-electrode patterning (Methods).

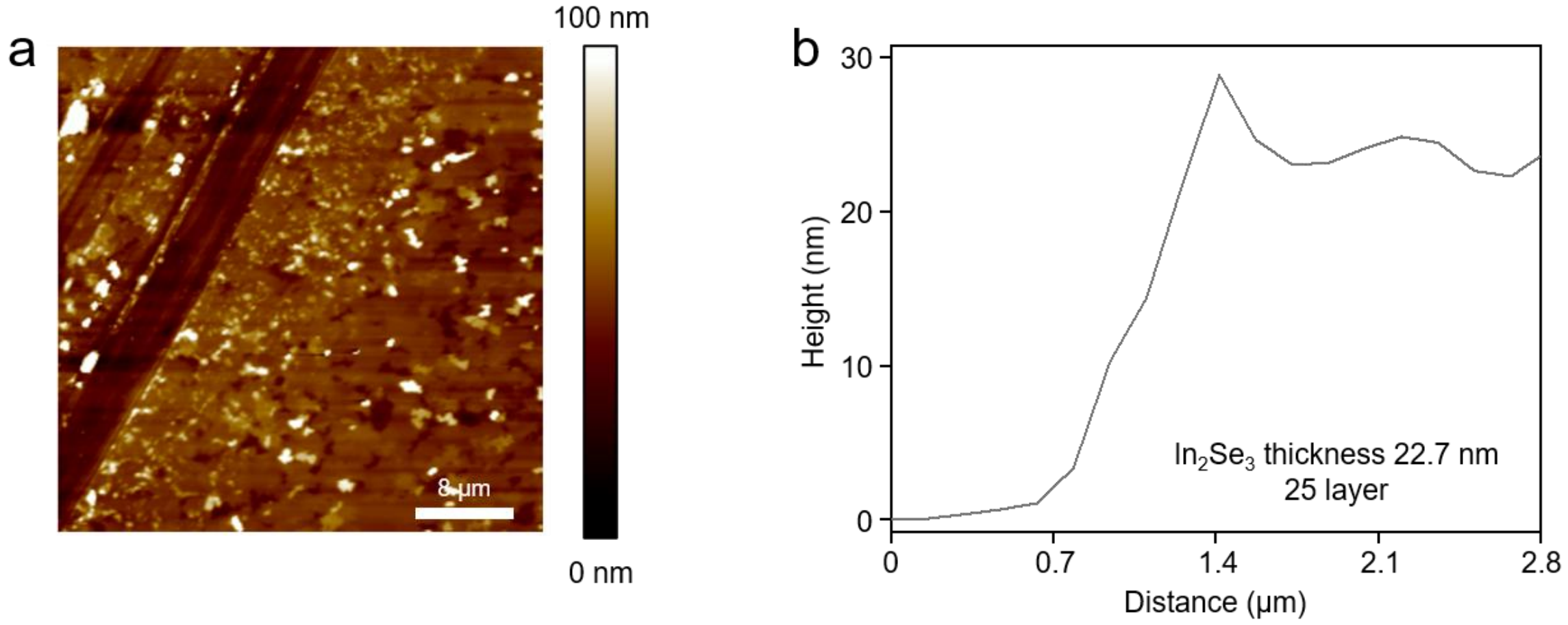

**Supplementary Figure 13.** (a) AFM topographic image of the $In_2Se_3$ film used for the PFM measurements in Figure 2. The scratch was intentionally introduced to determine the film thickness. (b) Height profile of the $In_2Se_3$ film measured across the scratched region.

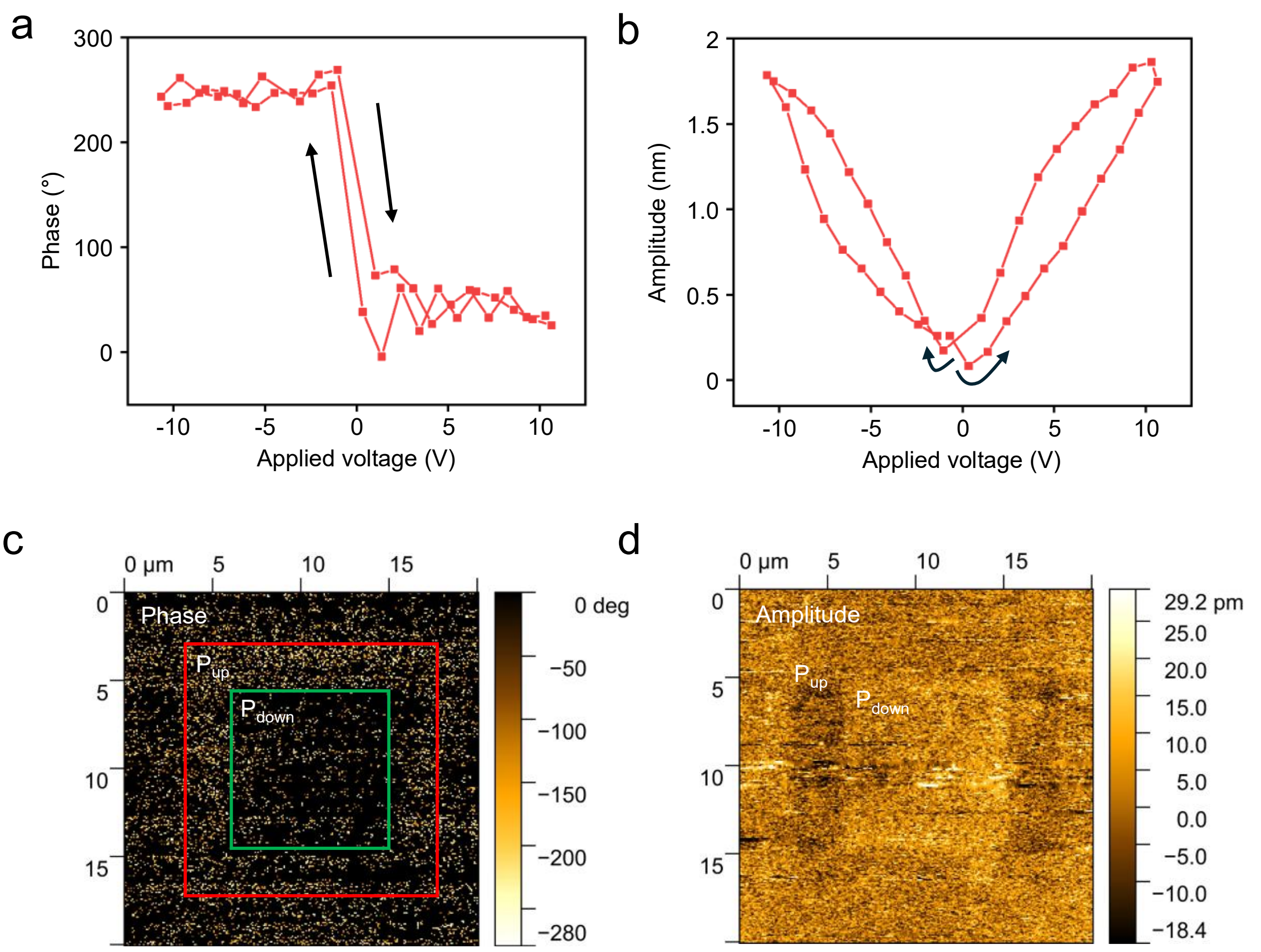

**Supplementary Figure 14.** Piezoresponse force microscopy (PFM) characterization of β-$In_2Se_3$. (a) PFM phase hysteresis loop and (b) corresponding amplitude butterfly loop measured as a function of applied voltage. (c, d) Corresponding PFM phase (c) and amplitude (d) images acquired after sequential electrical poling of nested square regions with opposite polarization directions. The red and green squares in (c) indicate the outer upward-polarized ($P_{up}$) and inner downward-polarized ($P_{down}$) regions, respectively. The domains were written on a ~5-nm-thick film with tip biases of ±10 V, and the images were acquired about 1 h after writing.

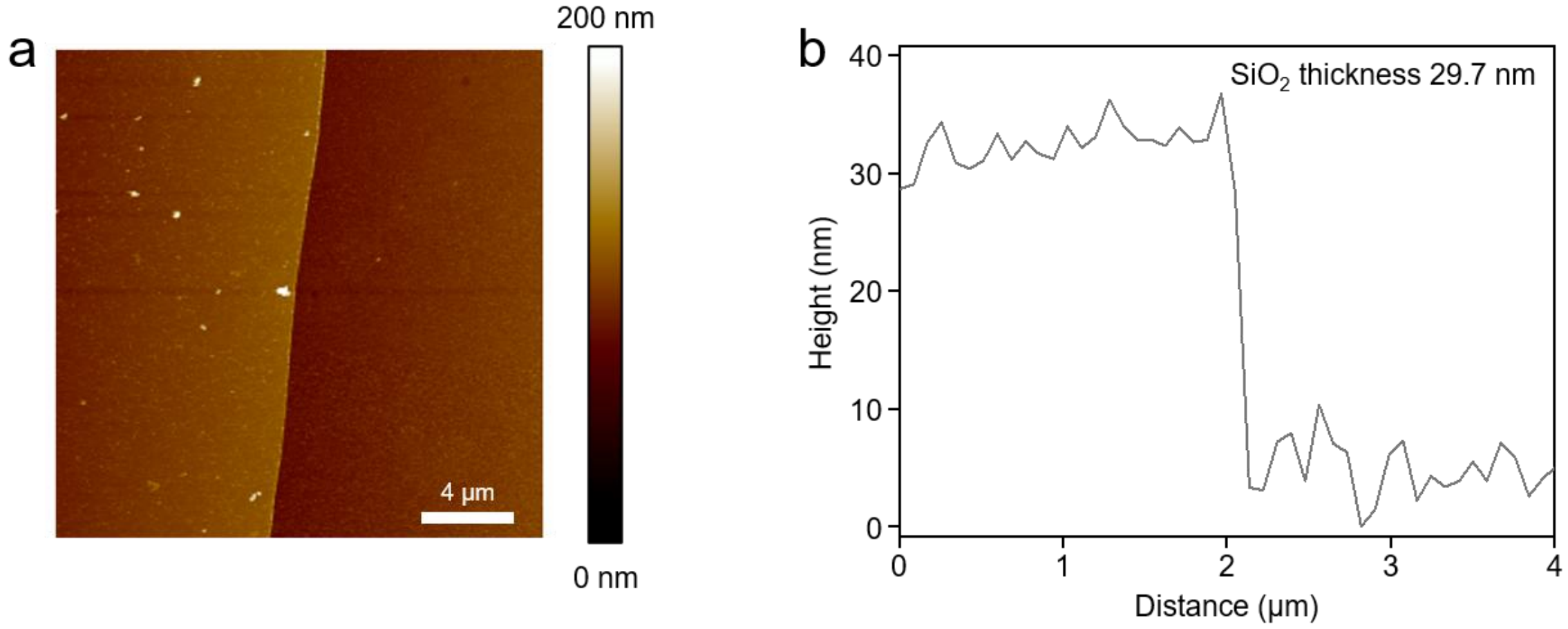

**Supplementary Figure 15.** (a) AFM topographic image of the $SiO_2$ film deposited on a Si substrate placed alongside the device substrate during electron-beam evaporation. The Si substrate was used as a reference for determining the thickness of the $SiO_2$ layer employed in the capacitor. (b) Corresponding height profile of the $SiO_2$ film.

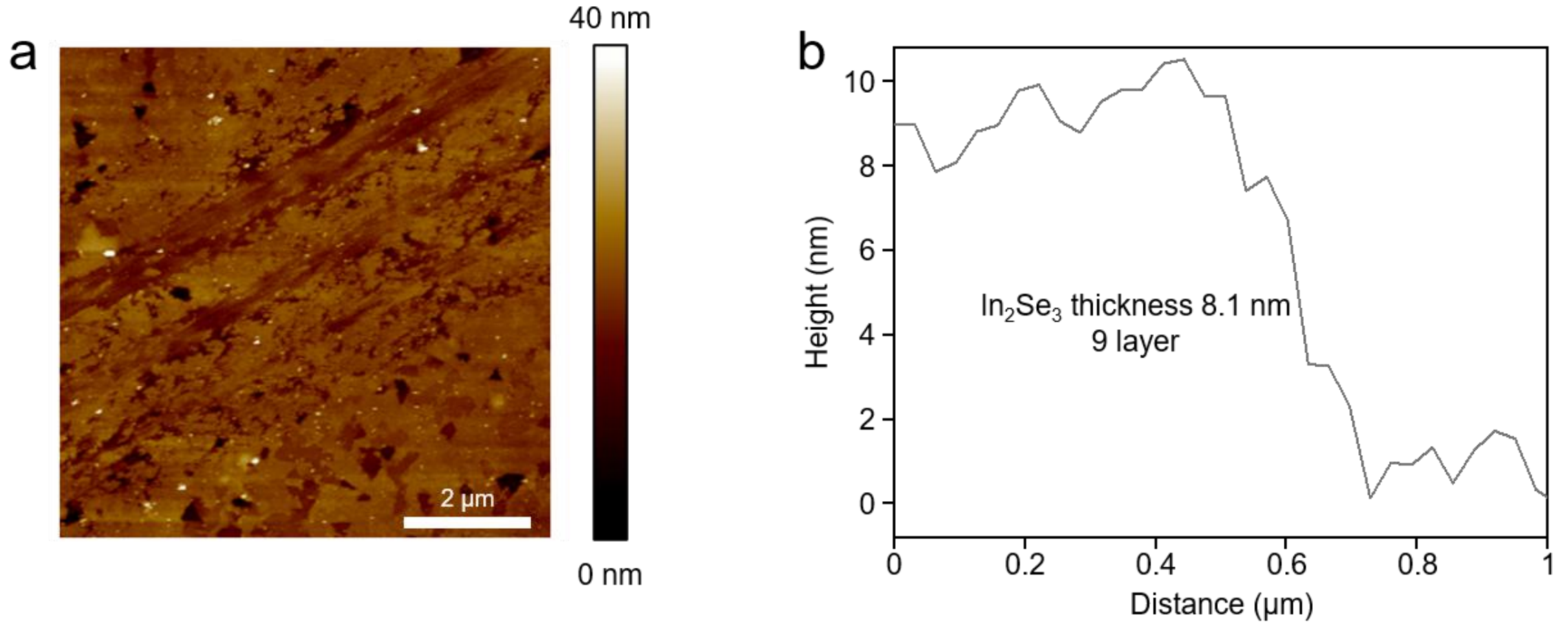

**Supplementary Figure 16.** (a) AFM topographic image of the $In_2Se_3$ film in the capacitor used in Figure 2. The scratch was intentionally introduced to determine the film thickness. (b) Height profile of the $In_2Se_3$ film across the scratched region.

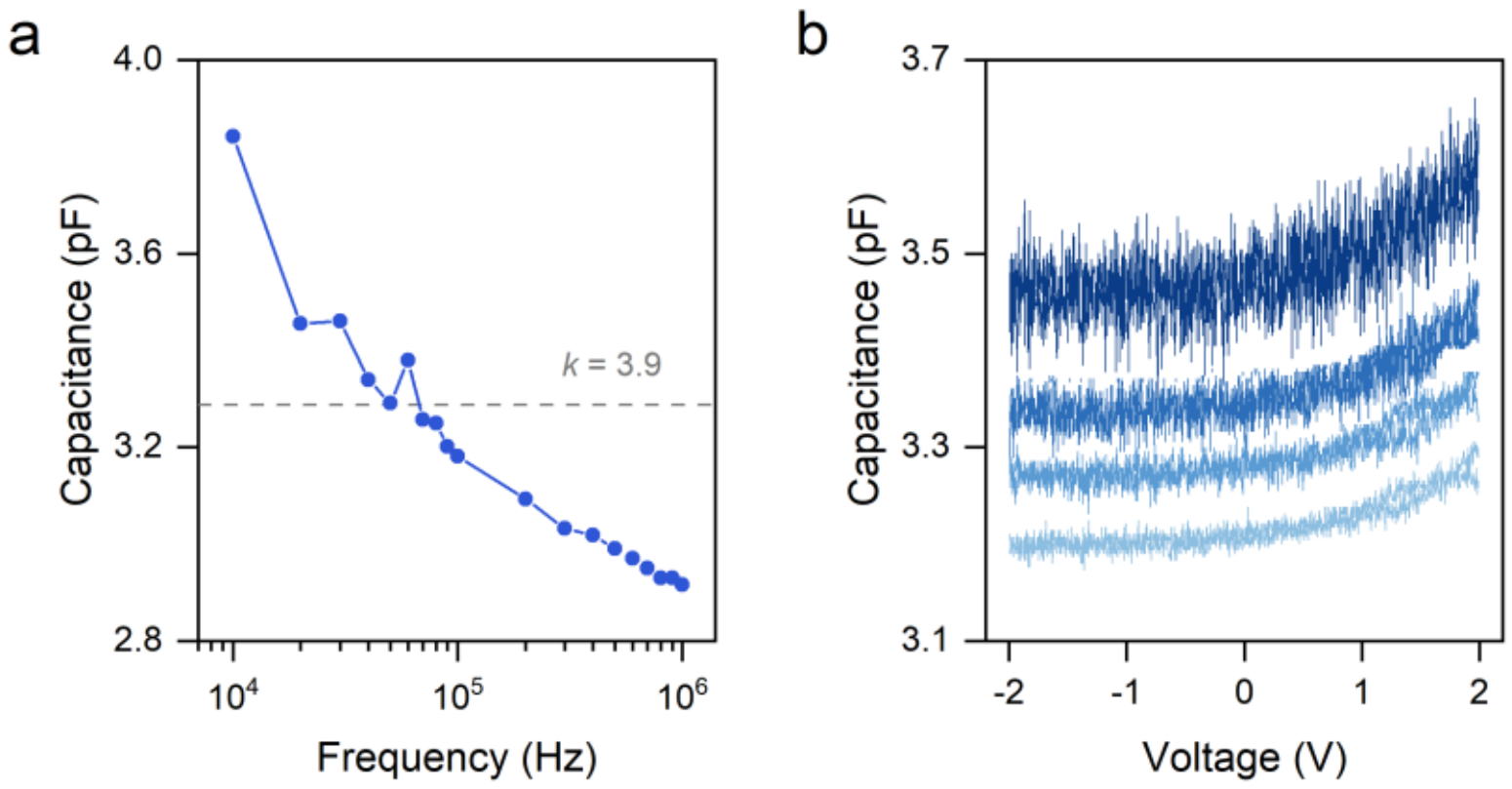

**Supplementary Figure 17.** Ti/Au/$SiO_2$/Au reference capacitor without β-$In_2Se_3$. The $SiO_2$ layer (29.7 nm) was deposited with the same method and conditions as in the β-$In_2Se_3$ capacitors, and the top electrode has a radius of 30 µm. (a) Capacitance–frequency characteristics from 10 kHz to 1 MHz. The capacitance of 3.29 pF at 50 kHz corresponds to *k* = 3.9 (dashed line). (b) Capacitance–voltage characteristics from four measurements over ±2 V. No peaks are observed, and the capacitance changes by about 3% over this range.

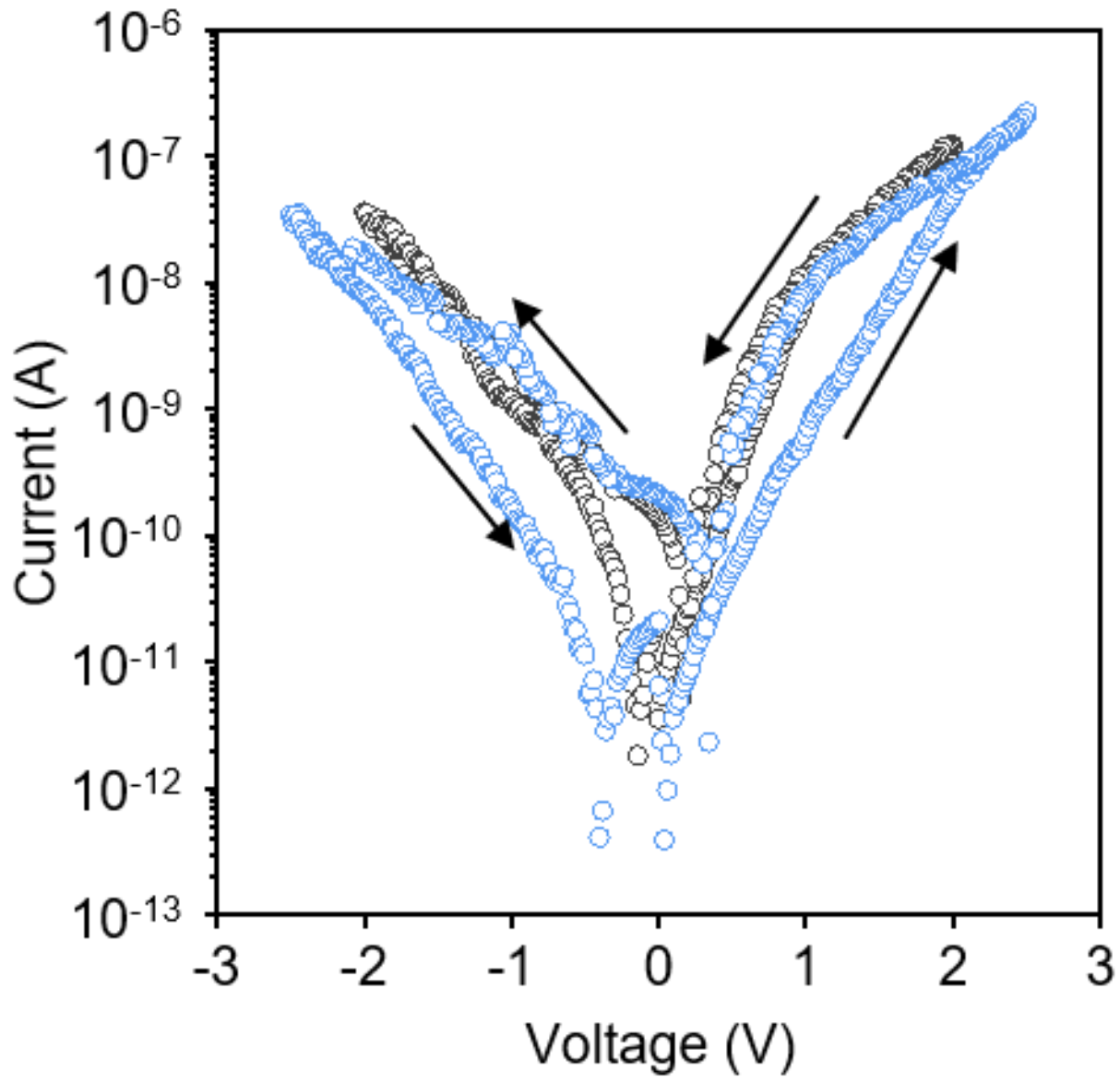

**Supplementary Figure 18.** DC I–V characteristics of the capacitor shown in Figure 2 over ±2.5 V, below the switching voltage, showing hysteretic leakage current. This hysteretic leakage corresponds to the ferroelectric-diode (ferrodiode) effect reported for AlScN capacitors, in which the leakage current depends on the polarization state[S1,S2]. The blue curve was measured over a wider voltage range than the gray curve.

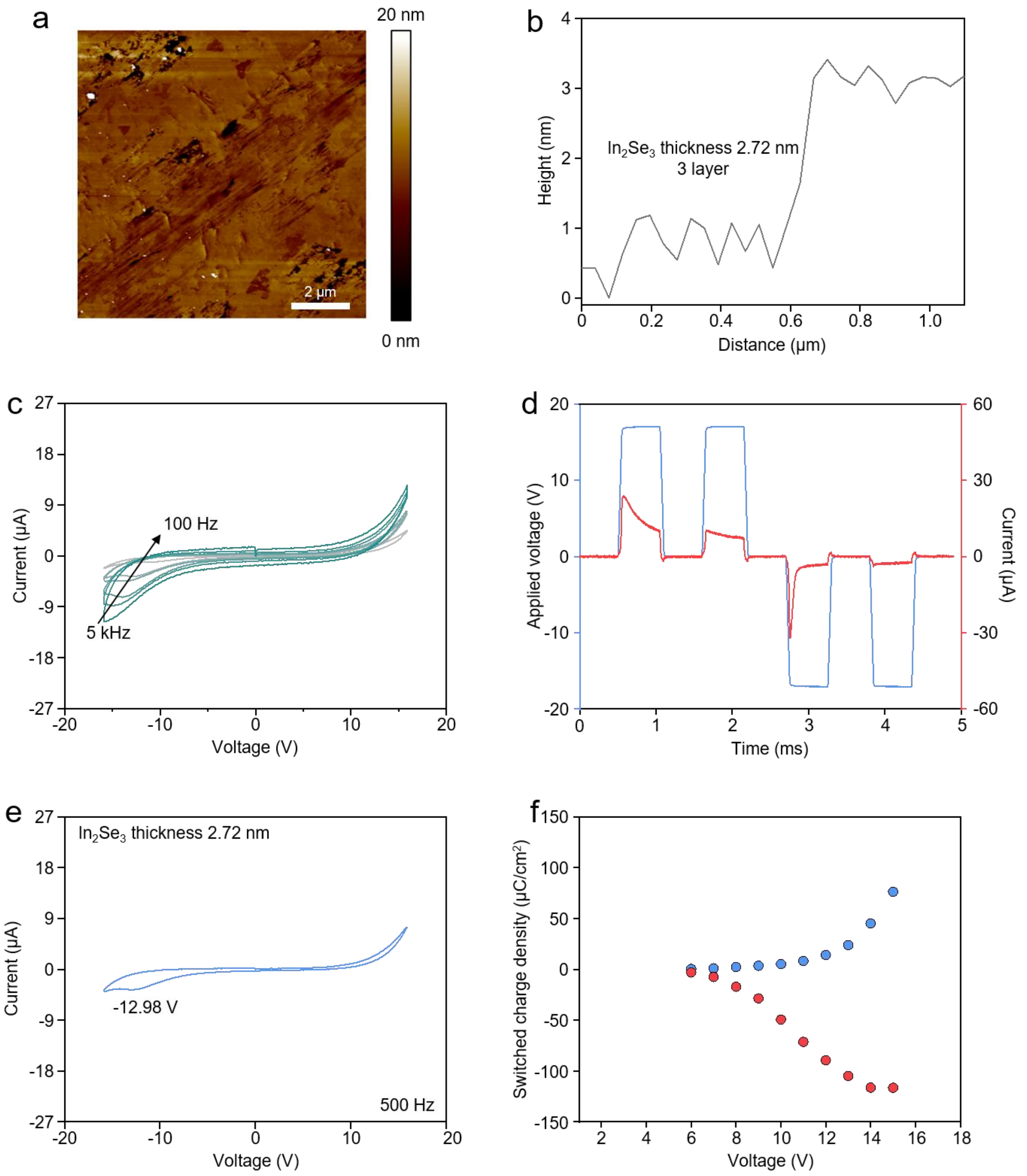

**Supplementary Figure 19.** (a) AFM topographic image of the 2.72-nm-thick $In_2Se_3$ film. The scratch was intentionally introduced for film-thickness measurement. (b) Corresponding height profile across the scratched region. (c) Current–voltage (I–V) hysteresis loops of the 2.72-nm-thick $In_2Se_3$ device measured at different voltage frequencies. (d) Time-resolved current response of the device under applied voltage pulses. (e) I–V hysteresis loop measured at 500 Hz for a 2.72-nm-thick $In_2Se_3$ device. (f) Switched charge density of the 2.72-nm-thick $In_2Se_3$ device as a function of pulse amplitude.

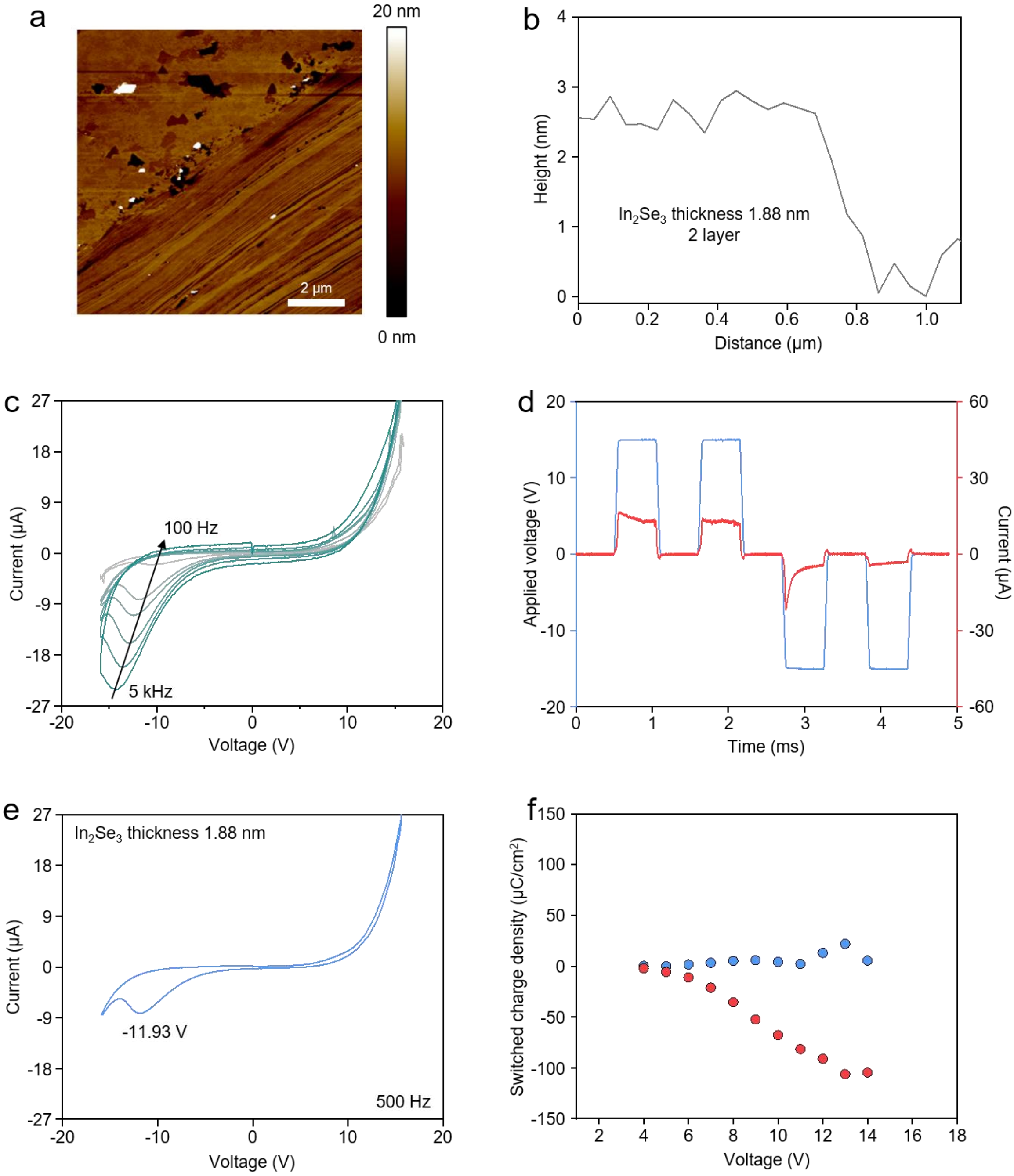

**Supplementary Figure 20.** (a) AFM topographic image of the 1.88-nm-thick $In_2Se_3$ film. The scratch was intentionally introduced for film-thickness measurement. (b) Corresponding height profile across the scratched region. (c) Current–voltage (I–V) hysteresis loops of the 1.88-nm-thick $In_2Se_3$ device measured at different voltage frequencies. (d) Time-resolved current response of the device under applied voltage pulses. (e) I–V hysteresis loop measured at 500 Hz for a 1.88-nm-thick $In_2Se_3$ device. (f) Switched charge density of the 1.88-nm-thick $In_2Se_3$ device as a function of pulse amplitude.

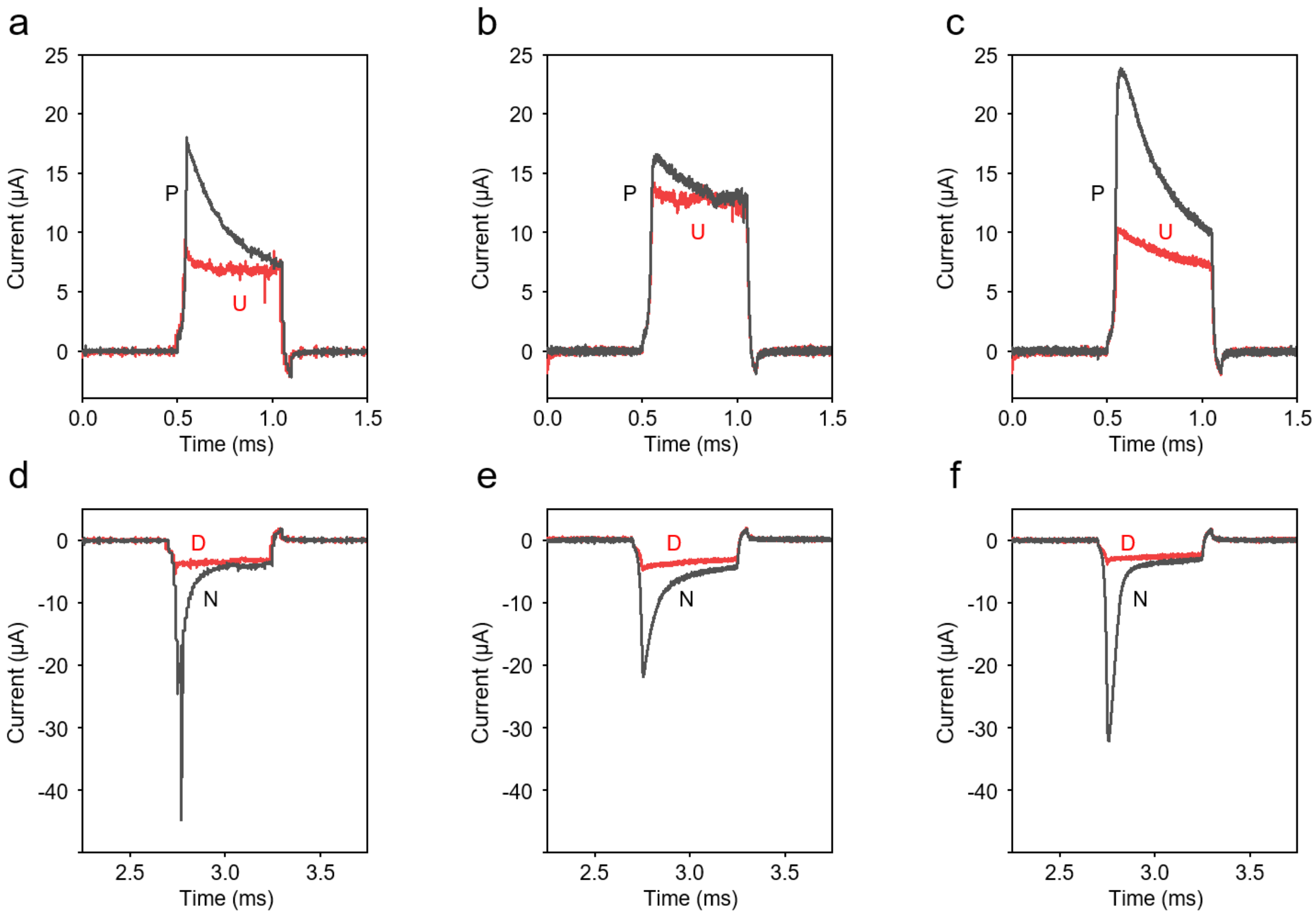

**Supplementary Figure 21.** Overlaid current transients from PUND measurements of β-$In_2Se_3$ capacitors. (a-c) Current responses during the positive switching (P) and subsequent nonswitching (U) pulses for (a) the capacitor in Figure 2, (b) the 1.88-nm-thick β-$In_2Se_3$ capacitor, and (c) the 2.72-nm-thick β-$In_2Se_3$ capacitor. (d-f) Corresponding current responses during the negative switching (N) and subsequent nonswitching (D) pulses for the devices shown in (a-c), respectively. Black and red curves represent the switching (P or N) and nonswitching (U or D) pulses, respectively. The overlaid transients allow direct comparison of consecutive pulse responses used for PUND subtraction.

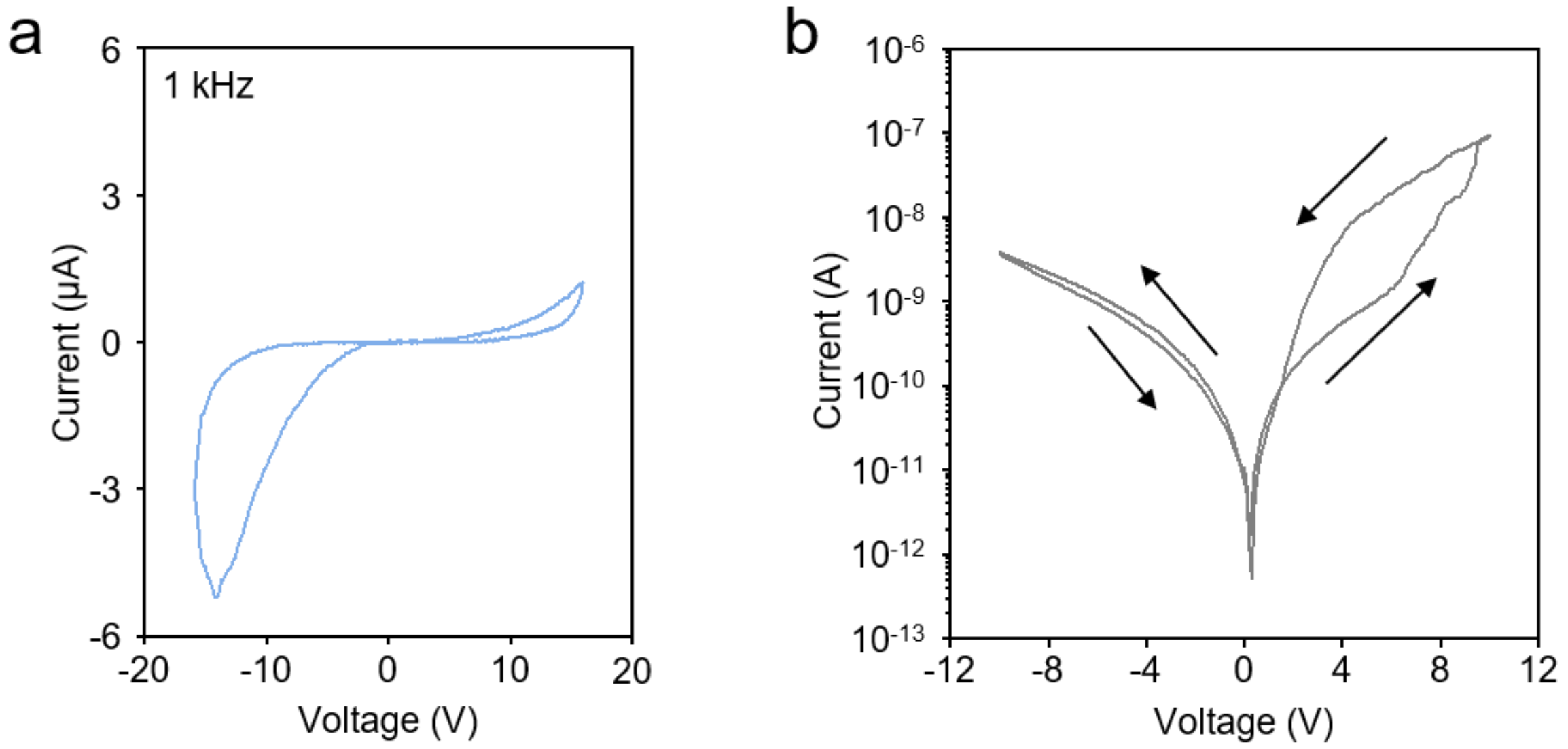

**Supplementary Figure 22.** Hysteretic I–V characteristics of the lateral device in Figure 3c (electrode gap, 500 nm). (a) Current–voltage (I–V) hysteresis loops measured at a voltage frequency of 1 kHz. (b) DC I–V characteristics of the device, showing clear hysteretic behavior, which corresponds to the ferrodiode effect[S1,S2] noted in Supplementary Figure 18.

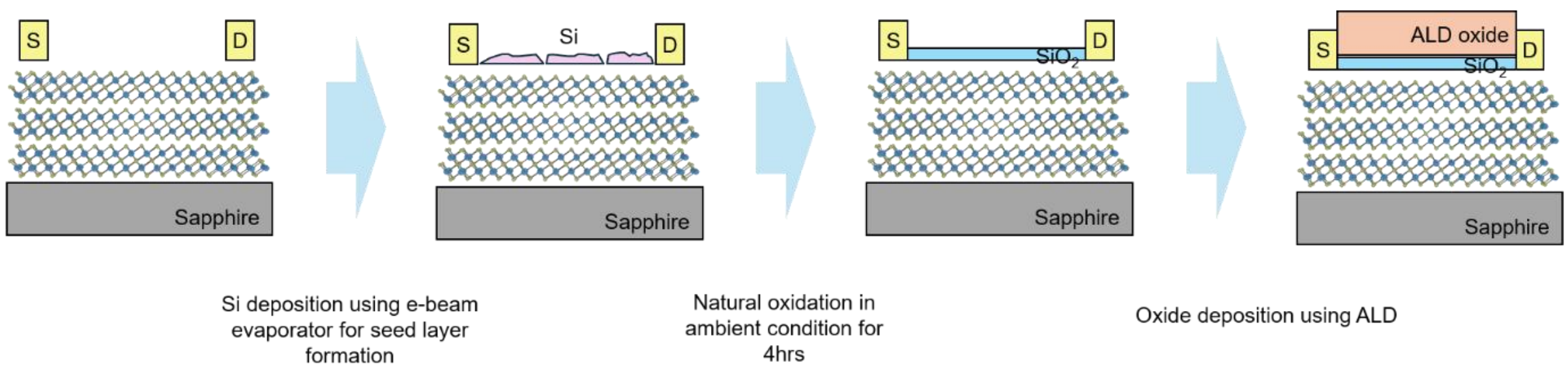

**Supplementary Figure 23.** Fabrication process of the $In_2Se_3$ FeS-FET device.

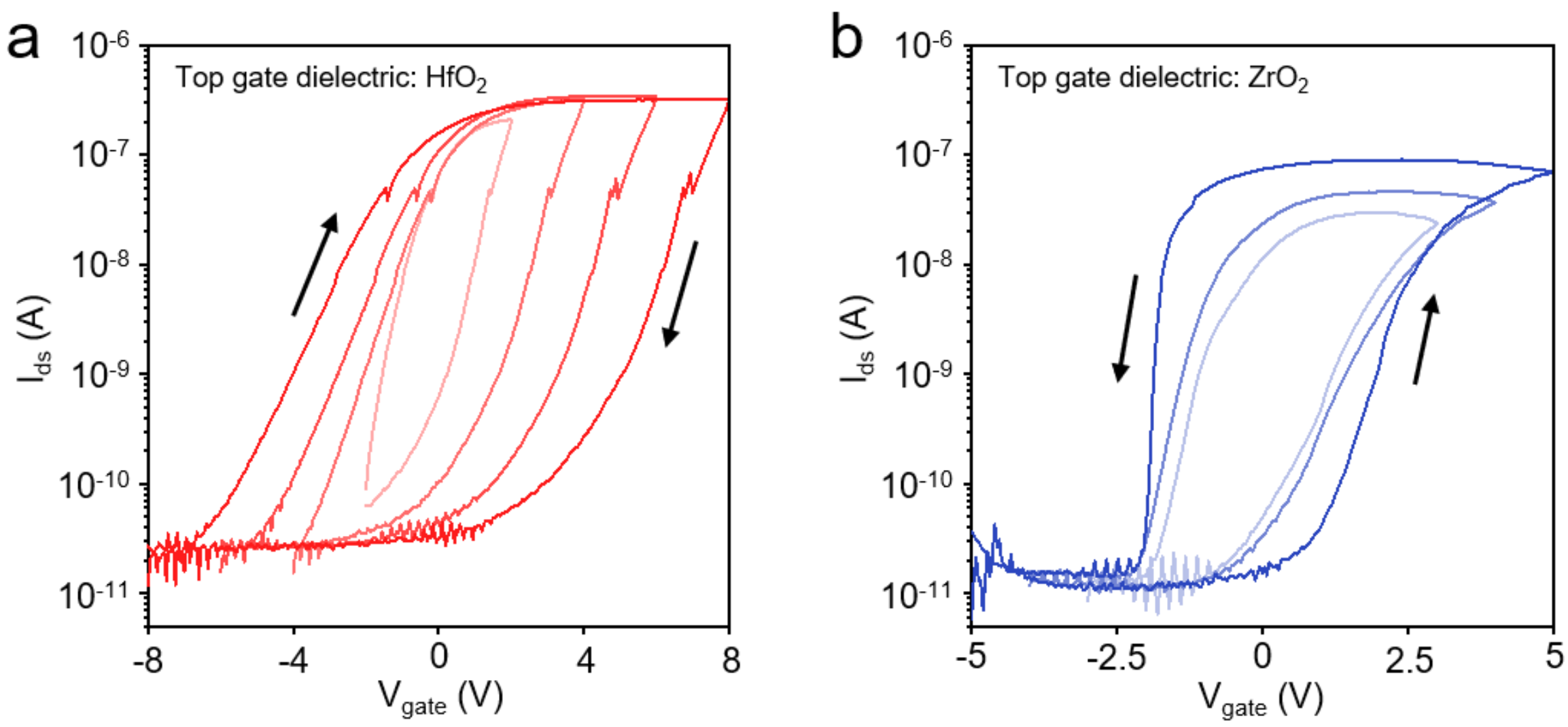

**Supplementary Figure 24.** Transfer characteristics of the FeS-FET device measured over increasing gate-voltage ranges, (a) using $HfO_2$ and (b) using $ZrO_2$ as the gate dielectric; the memory window widens with the sweep range in both cases (Supplementary Tables 1 and 2).

**Supplementary Table 1.** Memory window, on/off current ratio, and subthreshold swing of the $HfO_2$-gated FeS-FET as a function of the gate-voltage sweep range.

| $V_g$ sweep range (V) | ± 2 V | ± 4 V | ± 6 V | ± 8 V |
|---|---|---|---|---|
| Memory window (V) | 1.89 | 3.77 | 6.26 | 9.02 |
| On/Off | 3345.37 | 21074.66 | 22095.79 | 27768.19 |
| Subthreshold swing (mV/dec) | 604.61 | 1012.27 | 1290.00 | 1528.41 |

**Supplementary Table 2.** Memory window, on/off current ratio, and subthreshold swing of the $ZrO_2$-gated FeS-FET as a function of the gate-voltage sweep range.

| $V_g$ sweep range (V) | ± 3V | ± 4 V | ± 5 V |
|---|---|---|---|
| Memory window (V) | 2.43 | 3.23 | 4.31 |
| On/Off | 2884.87 | 4598.56 | 8574.76 |
| Subthreshold swing (mV/dec) | 1034.95 | 1081.48 | 760.31 |

* The subthreshold swing (SS) was calculated as the average over two decades of drain current, from $10^{-10}$ to $10^{-8}$ A.

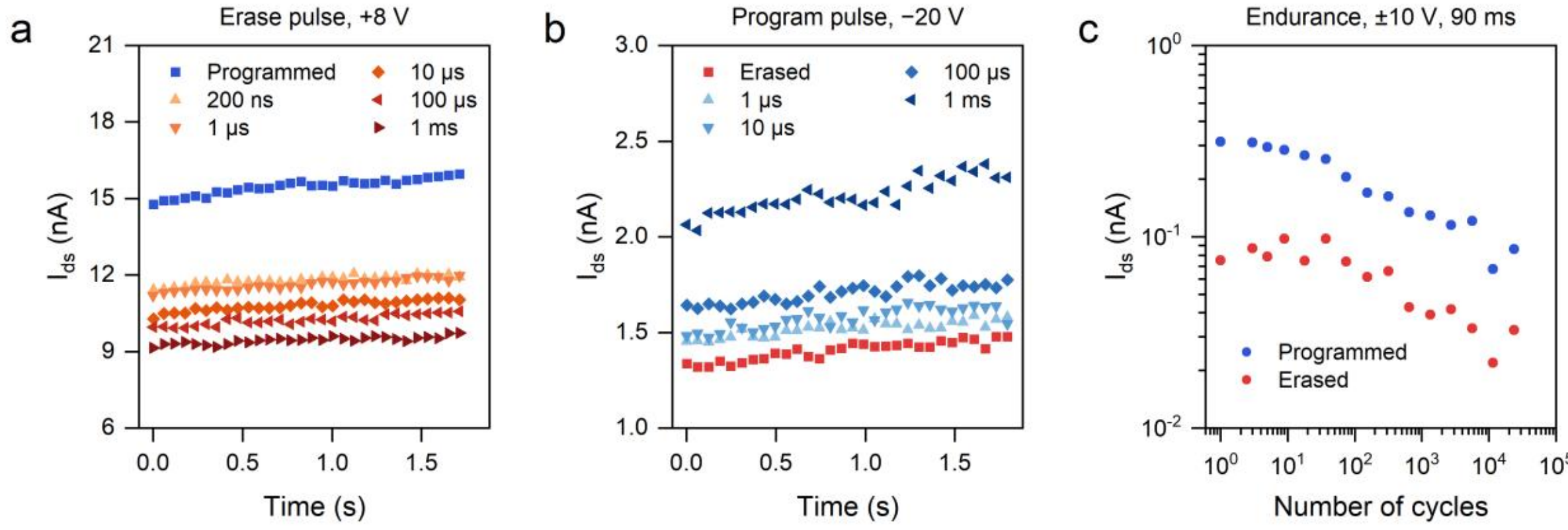

**Supplementary Figure 25.** Pulse response of the $HfO_2$-gated FeS-FET. (a) Erase: after programming at −9 V for 900 ms, erase pulses of +8 V with widths from 200 ns to 1 ms were applied, and $I_{ds}$ was then read as a function of time. Relative to the programmed state, $I_{ds}$ decreased by 24, 25, 30, 34 and 39% for 200 ns, 1 µs, 10 µs, 100 µs and 1 ms, respectively. (b) Program: after erasing at +9 V for 900 ms, program pulses of −20 V with widths from 1 µs to 1 ms were applied. Relative to the erased state, $I_{ds}$ increased by 8, 12, 22 and 58% for 1 µs, 10 µs, 100 µs and 1 ms, respectively. (c) Endurance with ±10 V, 90-ms program and erase pulses. The ratio of the programmed to the erased current remained at 2.5–4 up to $2.4 \times 10^4$ cycles while both currents decreased. The endurance was limited by degradation of the $HfO_2$ gate dielectric, and the next point ($4.9 \times 10^4$ cycles), at which the two currents were no longer distinguishable, is not shown. All currents were read at $V_{gate} = 0$ V and $V_{ds} = 1$ V. Pulses shorter than 200 ns are not included, because the applied voltage did not reach the set value for pulse widths of 100 ns or less.

**Supplementary Table 3.** Polymorphs of $In_2Se_3$ relevant to ferroelectricity.

| Polymorph | In coordination | Thermal stability | Polar order | Vertical switching quantified ($P_r$, $E_c$) | Stacking degree of freedom | Large-area growth | Refs. |
|---|---|---|---|---|---|---|---|
| α (ZB′ / WZ′) | tetrahedral + octahedral | transforms to β at ~200–270 °C | out-of-plane + in-plane (dipole-locked) | PFM and tunnel junctions; $P_r$ from PUND in flakes | ZB′ versus WZ′ stacking | CVD/MBE; WZ′ reported once | 12–14, 25, 35, S3, S4, S5, S6 |
| β′ | octahedral | stable below ~200–250 °C, converts to β on heating | in-plane only; nanostripe (anti/ferri)electric | none (no out-of-plane component) | — | exfoliated / MBE | S7, S8, S9 |
| β | octahedral | high-temperature polymorph, retained on cooling in epitaxial films; obtainable at 400 °C | nominally none (centrosymmetric); this work: out-of-plane 1.9 and in-plane 21 μC/cm$^2$ (calculated) | this work: $E_c$ = 0.9–1.0 MV/cm (k = 17), apparent $P_r$ = 11–20 μC/cm$^2$ (PUND at ~1.0 MV/cm), PFM, STEM | | MOCVD 2-inch (this work); CVD cm-scale (14); 400 °C deposition (S10) | 13, 14, S10, this work |

**Supplementary Table 4.** Benchmark of β-$In_2Se_3$ against representative 2D and 3D ferroelectrics. Ten-year retention values are extrapolations; all other values are measured. Plain numbers refer to the main-text reference list and S-numbers to the Supplementary References below.

| **Material** | **Growth / area** | **Thinnest ferroelectric layer** | **Vertical $E_c$ (MV/cm)** | **Measured $P_r$ (μC/cm$^2$)** | **Endurance (cycles)** | **Retention** | **Ferroelectric semiconductor channel** | **Device statistics** | **Refs.** |
|---|---|---|---|---|---|---|---|---|---|
| β-$In_2Se_3$ (this work) | MOCVD, 2-inch wafer | 1.9 nm (bilayer) | 0.9–1.0 (k = 17), constant from 1.9 to 8 nm | 11–20 (apparent $P_r$, PUND at ~1.0 MV/cm) | $10^7$ (adaptive voltage); 6 × $10^5$ (fixed 12 V) | 1.3 × $10^5$ s (negative state); $10^4$ s (positive state) | Yes (FeS-FET) | 100 devices, CV 4.3% | this work |
| α-$In_2Se_3$ | CVD/MBE flakes; large-area films reported | few nm | ≈4 (6 nm); rises below ~25 nm | ≈1–7 (flakes) | $10^3$–$10^5$ | $10^3$–3 × $10^4$ s | Yes (FeS-FET) | none | 8, 14, 35, 52, S4, S5, S6, S11, S12, S13 |
| β′-$In_2Se_3$ | exfoliated / MBE | — | no vertical switching reported | in-plane only (PFM) | — | — | — | none | S7, S8, S9 |
| γ-InSe (sliding) | exfoliated bulk | few-layer | — | small (PFM only) | — | — | — | none | 11, S14 |
| 3R-$MoS_2$ (sliding) | exfoliated / stacked | bilayer | ≈12 | ≈0.5 | $10^{11}$ | $10^6$ s | FET reported | none | 33, 34, 62, S15, S16 |
| Parallel bilayer hBN | exfoliated / stacked | bilayer | ≈3 | <1 | $10^{11}$ | $10^6$ s | — | none | 33, S17, S18 |
| $CuInP_2S_6$ | exfoliated | ≈4 nm | — | ≈8 (20 nm) | $10^3$ | $10^4$ s | — | none | 53, S19 |
| HZO | ALD, wafer | 1–2 nm | 1–5 (rises below ~5 nm) | 10–25 | $10^9$–$10^{12}$ | 10 yr (extrapolated) | No | yes | 36, 37, 40, 44 |
| AlScN | sputtering / MBE, wafer | 5 nm | 4–6 | 20–135 | $10^{10}$ | 10 yr (extrapolated) | No | yes | 29, 48–51 |
| PZT / $BaTiO_3$ | PLD / MOCVD, wafer | ≈5 nm | 0.1–0.5 | 20–40 | $10^9$ | 10 yr (extrapolated) | No | yes | 45, 47, 60, 61 |

## Supplementary References

S1 Liu, X. *et al.* Aluminum scandium nitride-based metal–ferroelectric–metal diode memory devices with high on/off ratios. *Applied Physics Letters* **118**, 202901 (2021). https://doi.org/10.1063/5.0051940

S2 Liu, X. *et al.* Reconfigurable compute-in-memory on field-programmable ferroelectric diodes. *Nano Letters* **22**, 7690-7698 (2022). https://doi.org/10.1021/acs.nanolett.2c03169

S3 Jiang, Y. *et al.* 2D ferroelectric narrow-bandgap semiconductor wurtzite′-type α-In2Se3 and its silicon-compatible growth. *Nature Communications* **16**, 7364 (2025). https://doi.org/10.1038/s41467-025-62822-7

S4 Zhou, Y. *et al.* Out-of-plane piezoelectricity and ferroelectricity in layered α-In2Se3 nanoflakes. *Nano Letters* **17**, 5508-5513 (2017). https://doi.org/10.1021/acs.nanolett.7b02198

S5 Xiao, J. *et al.* Intrinsic two-dimensional ferroelectricity with dipole locking. *Physical Review Letters* **120**, 227601 (2018). https://doi.org/10.1103/PhysRevLett.120.227601

S6 Cui, C. *et al.* Intercorrelated in-plane and out-of-plane ferroelectricity in ultrathin two-dimensional layered semiconductor In2Se3. *Nano Letters* **18**, 1253-1258 (2018). https://doi.org/10.1021/acs.nanolett.7b04852

S7 Zheng, C. *et al.* Room temperature in-plane ferroelectricity in van der Waals In2Se3. *Science Advances* **4**, eaar7720 (2018). https://doi.org/10.1126/sciadv.aar7720

S8 Xu, C. *et al.* Two-dimensional antiferroelectricity in nanostripe-ordered In2Se3. *Physical Review Letters* **125**, 047601 (2020). https://doi.org/10.1103/PhysRevLett.125.047601

S9 Liu, D. S. H. *et al.* Evidence for in-plane electrical polarization in 3R-β′-In2Se3 thin films grown by molecular beam epitaxy. *ACS Applied Materials & Interfaces* **17**, 35661-35672 (2025). https://doi.org/10.1021/acsami.5c01185

S10 Lee, S. *et al.* Low-temperature processed beta-phase In2Se3 ferroelectric semiconductor thin film transistors. *2D Materials* **9**, 025023 (2022). https://doi.org/10.1088/2053-1583/ac5b17

S11 Park, S., Lee, D., Kang, J., Choi, H. & Park, J.-H. Laterally gated ferroelectric field effect transistor (LG-FeFET) using α-In2Se3 for stacked in-memory computing array. *Nature Communications* **14**, 6778 (2023). https://doi.org/10.1038/s41467-023-41991-3

S12 Li, Y. *et al.* Large-area polymorphic In2Se3 ferroelectric transistor array for stable nonvolatile storage and high-precision neuromorphic computing. *Small*, e03908 (2025). https://doi.org/10.1002/smll.202503908

S13 Io, W. F. *et al.* Temperature- and thickness-dependence of robust out-of-plane ferroelectricity in CVD grown ultrathin van der Waals α-In2Se3 layers. *Nano Research* **13**, 1897-1902 (2020). https://doi.org/10.1007/s12274-020-2640-0

S14 Sui, F. *et al.* Atomic-level polarization reversal in sliding ferroelectric semiconductors. *Nature Communications* **15**, 3799 (2024). https://doi.org/10.1038/s41467-024-48218-z

S15 Wang, X. *et al.* Interfacial ferroelectricity in rhombohedral-stacked bilayer transition metal dichalcogenides. *Nature Nanotechnology* **17**, 367-371 (2022). https://doi.org/10.1038/s41565-021-01059-z

S16 Weston, A. *et al.* Interfacial ferroelectricity in marginally twisted 2D semiconductors. *Nature Nanotechnology* **17**, 390-395 (2022). https://doi.org/10.1038/s41565-022-01072-w

S17 Yasuda, K., Wang, X., Watanabe, K., Taniguchi, T. & Jarillo-Herrero, P. Stacking-

engineered ferroelectricity in bilayer boron nitride. *Science* **372**, 1458-1462 (2021). https://doi.org/10.1126/science.abd3230
S18 Vizner Stern, M. *et al.* Interfacial ferroelectricity by van der Waals sliding. *Science* **372**, 1462-1466 (2021). https://doi.org/10.1126/science.abe8177
S19 Liu, F. *et al.* Room-temperature ferroelectricity in CuInP2S6 ultrathin flakes. *Nature Communications* **7**, 12357 (2016). https://doi.org/10.1038/ncomms12357